\documentclass[aps,pra, reprint,superscriptaddress,a4paper,longbibliography]{revtex4-1}
\usepackage{amsmath,amssymb}
\usepackage[utf8]{inputenc}
\usepackage{graphicx}
\usepackage{xcolor}
\usepackage{hyperref}
\hypersetup{colorlinks=true,citecolor={blue},linkcolor={blue},urlcolor={blue}}
\usepackage{braket}
\usepackage{bm}
\usepackage{enumitem}
\usepackage[pass]{geometry}
\usepackage{newunicodechar}
\usepackage{notes2bib}

\usepackage[separate-uncertainty=true,multi-part-units=single,group-separator={,}]{siunitx}
\DeclareSIUnit[per-mode=symbol,per-symbol=p]{\ueV}{\micro\electronvolt}
\DeclareSIUnit[per-mode=symbol,per-symbol=p]{\um}{\micro\meter}

\newcommand{\capind}[1]{({#1})}
\newcommand{\pth}{P_{\text{th}}}

\graphicspath{{./figs/}}
\usepackage{float}

\begin{document}

\author{H. Ohadi}
\email{ho35@st-andrews.ac.uk}
\affiliation{Department of Physics, Cavendish Laboratory, University of Cambridge, Cambridge CB3 0HE, United Kingdom}
\affiliation{SUPA, School of Physics and Astronomy, University of St Andrews, St Andrews, KY16 9SS, United Kingdom}
\author{Y. del Valle-Inclan Redondo}
\affiliation{Department of Physics, Cavendish Laboratory, University of Cambridge, Cambridge CB3 0HE, United Kingdom}
\author{A. J. Ramsay}
\affiliation{Hitachi Cambridge Laboratory, Hitachi Europe Ltd., Cambridge CB3 0HE, UK}
\author{Z. Hatzopoulos}
\affiliation{FORTH, Institute of Electronic Structure and Laser, 71110 Heraklion, Crete, Greece}
\author{T. C. H. Liew}
\affiliation{School of Physical and Mathematical Sciences, Nanyang Technological University 637371, Singapore}
\author{P. R. Eastham}
\affiliation{School of Physics and CRANN, Trinity College Dublin, Dublin 2, Ireland}
\author{P. G. Savvidis}
\affiliation{FORTH, Institute of Electronic Structure and Laser, 71110 Heraklion, Crete, Greece}
\affiliation{ITMO University, St.\ Petersburg 197101, Russia}
\affiliation{Department of Materials Science and Technology, University of Crete, 71003 Heraklion, Crete, Greece}
\author{J. J. Baumberg}
\email{jjb12@cam.ac.uk}
\affiliation{Department of Physics, Cavendish Laboratory, University of Cambridge, Cambridge CB3 0HE, United Kingdom}

\begin{abstract} We demonstrate that the synchronization of a lattice of
solid-state condensates when inter-site tunnelling is switched on, depends
strongly on the weak local disorder. This finding is vital for implementation of
condensate arrays as computation devices. The condensates here are nonlinear
bosonic fluids of exciton-polaritons trapped in a weakly disordered Bose-Hubbard
potential, where the nearest neighboring tunneling rate (Josephson coupling) can
be dynamically tuned. The system can thus be tuned from a localized to a
delocalized fluid as the number density, or the Josephson coupling between
nearest neighbors increases. The localized fluid is observed as a lattice of
unsynchronized condensates emitting at different energies set by the disorder
potential. In the delocalized phase the condensates synchronize, and long-range
order appears, evidenced by narrowing of momentum and energy distributions, new
diffraction peaks in momentum space, and spatial coherence between condensates.
Our work identifies similarities and differences of this nonequilibrium
crossover to the traditional Bose-glass to superfluid transition in atomic
condensates.  \end{abstract}

\title{Synchronization crossover of polariton condensates in weakly disordered lattices}

\maketitle


\section{Introduction}
\begin{figure*}
		\centering
		\includegraphics[width=.9\linewidth]{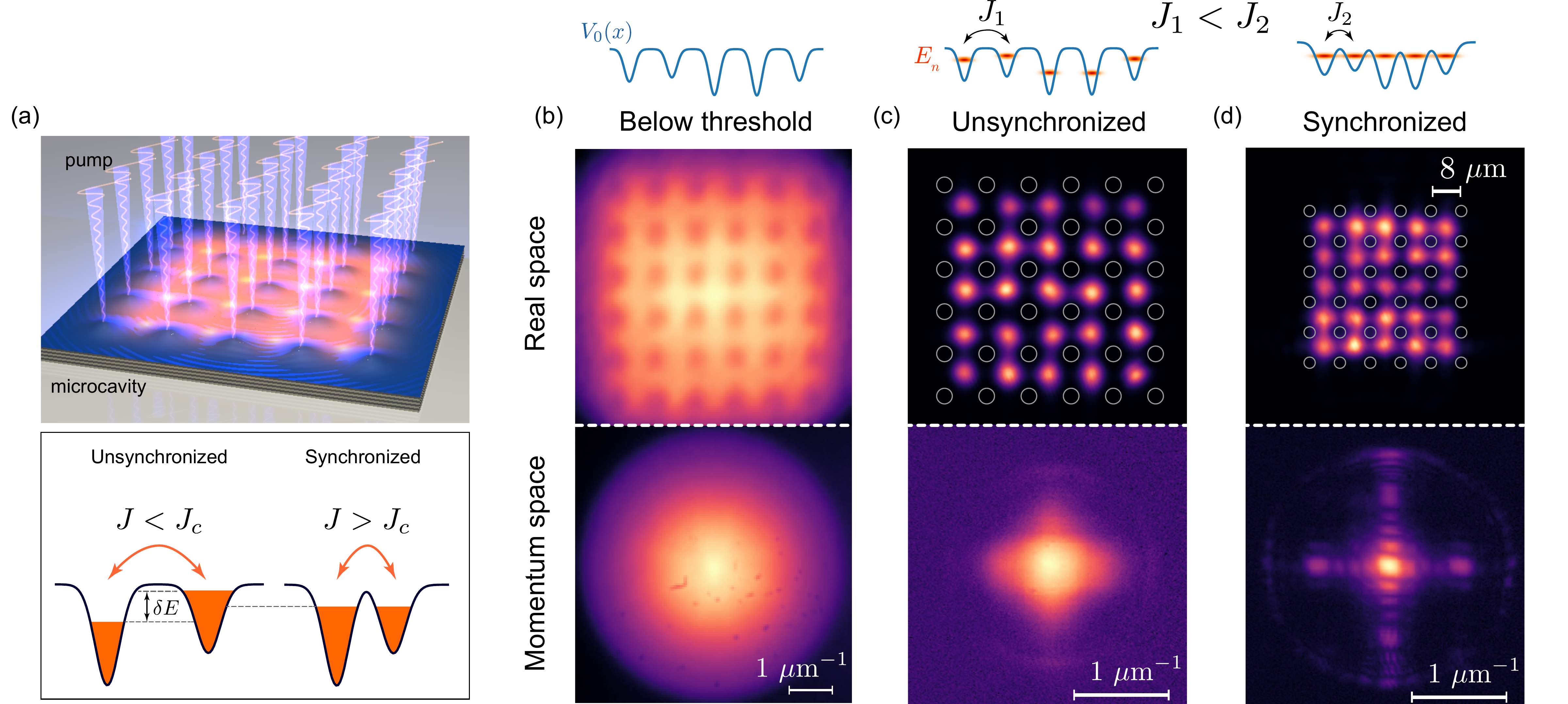}
		\caption{Below threshold, unsynchronized and synchronized phases. \capind{a}
		Schematic of condensation in optical lattice potentials (top panel), and
		synchronization of two energy-detuned condensates
		in a double-well potential (bottom panel). There is a nonzero energy
		detuning $\delta E$ at the UNSYNC phase but this vanishes at the
		SYNC phase as the Josephson coupling increase. \capind{b-d}
		Real-space and momentum space emission of \capind{b}
		uncondensed polaritons, \capind{c} unsynchronized (localized), and
		\capind{d} synchronized (delocalized) condensate lattice. Pump spots are
		marked by grey circles in \capind{c,d} and visible by dark spots in
		\capind{b}. The schematic diagrams above the panels show the trapping
		potential $V(x)$, Josephson coupling $J$ and energies of the condensates
		$E_n$. Momentum space images are in logarithmic scale.
}\label{fig:1}
\end{figure*}

In a lattice of trapped bosons, disorder inhibits coherent tunneling so
localizing condensation in real space. However on-site repulsive interactions
combined with a flow of bosons can tune neighboring condensates into resonance,
thus enhancing tunneling through the trap barriers. This subtle interplay is
crucial for understanding the rich phase diagram composed of Mott-insulator,
Bose-glass and superfluid, which acquires further complexity with pumping and
dissipation in nonequilibrium situations.  This situation is vital for
understanding the capabilities of lattices of condensates to act for instance as
quantum simulators of complex behavior. Exciton-polaritons (polaritons), which
are bosons composed of an admixture of quantum-well exciton and microcavity
photon~\cite{kavokin_microcavities_2007}, are an attractive system for studying
the nonequilibrium Bose-Hubbard system in two-dimensions
~\cite{carusotto_quantum_2013}. A lattice potential can be achieved either
structurally by depositing thin metallic films~\cite{lai_coherent_2007} and
etching~\cite{galbiati_polariton_2012} the microcavity or by photo-injecting
excitons with a spatially patterned laser~\cite{wertz_spontaneous_2010}. In the
latter case, the repulsive on-site interaction can be dynamically tuned through
polariton-polariton~\cite{takemura_polaritonic_2014} and polariton-exciton
nonlinearities~\cite{gao_polariton_2012}. For single polaritons the polariton
nonlinearity is weaker than the disorder potential, but when polaritons condense
into a macroscopic state~\cite{deng_condensation_2002,
kasprzak_bose-einstein_2006, baumberg_spontaneous_2008} the collective
nonlinearity can be large enough to exhibit effects such as
superfluidity~\cite{amo_superfluidity_2009, lerario_room-temperature_2017} and
solitons~\cite{amo_polariton_2011,sich_observation_2012,walker_dark_2017}.
Polaritons in one condensate `puddle' can tunnel out and drive neighboring
condensates~\cite{baas_synchronized_2008} described through the Josephson
mechanism~\cite{lagoudakis_coherent_2010}. The mechanism is more complicated
with disorder-induced energy detunings between two nearest neighbors, in the
case of repulsive nonlinearity~\cite{wouters_synchronized_2008}
(Fig.~\ref{fig:1}b). Here there is a critical Josephson tunneling ($J_c$), below
which an extended condensate separates into two localized condensates at
different energies, forming an unsynchronized (UNSYNC) phase. However above
$J_c$, the flow of polaritons from the higher-energy to the lower-energy
condensate collapses their energy detuning and synchronizes (SYNC) them due to
repulsive nonlinearities, as the higher energy condensate becomes less populated
and the lower energy condensate more populated than before (see lower panel in
Fig.~\ref{fig:1}a). The equivalent of a Bose-glass to superfluid crossover is
thus expected for a nonequilibrium polariton condensate lattice in a disordered
potential. This crossover occurs when the phase coherence length, which in a
disordered Bose insulator grows with increasing density or strength of the
Josephson
coupling,~\cite{malpuech_bose_2007,janot_superfluid_2013,nelsen_dissipationless_2013}
exceeds the overall size of the lattice. The Bose-glass to superfluid transition
was first observed in Josephson junction
arrays~\cite{geerligs_charging_1989,fisher_boson_1989} and it has recently been
observed in thermally equilibrated cold atom
lattices~\cite{meldgin_probing_2016}, but different features emerge for the
nonequilibrium lattice.

Here, we optically trap 25 polariton condensates inside a 5$\times$5 square
lattice within a planar semiconductor
microcavity~\cite{cristofolini_optical_2013,askitopoulos_polariton_2013}
(Fig.~\ref{fig:1}a). Optical trapping allows rapid and facile tuning of the
lattice constant as well as the nearest-neighbor coupling
strengths~\cite{ohadi_tunable_2016}. Due to residual weak disorder in the
microcavity, polaritons at different sites condense at slightly different
energies~\cite{ohadi_spin_2017}. When the lattice constant is large, the nearest
neighbor coupling is small and we observe an unsynchronized (UNSYNC) phase evidenced by a broad
peak in the momentum space corresponding to that of a single site condensate as well as a
distribution of condensate energies, and no long-range order. However, as the
nearest-neighbor coupling increases by reducing the lattice constant or
increasing the condensate number density (for higher excitation powers),
neighboring condensates collapse their energy detunings due to
polariton-polariton and polariton-exciton nonlinearities, until they all
synchronize (SYNC) resulting in an extended state with long-range order.  This appears
as progressive narrowing of the momentum space peak and appearance of sets of diffraction
peaks signifying a phase-locked fluid with a single delocalized macroscopic
wavefunction. Simulations show that disorder plays a key role in the crossover
width and the decay of long-range order.


\section{Formation of trapped condensates in a disordered potential}

We create polariton condensates by the nonresonant optical excitation of a
microcavity composed of GaAs quantum-wells sandwiched between two distributed
Bragg reflectors (DBR). The cavity top (bottom) DBR is made of 32 (35) pairs of
Al$_{0.15}$Ga$_{0.85}$As/AlAs layers of \SI{57.2}{\nm}/\SI{65.4}{\nm}. Four sets
of three \SI{10}{nm} GaAs quantum wells separated by \SI{10}{nm} thick layers of
Al$_{0.3}$Ga$_{0.7}$As are placed at the maxima of the cavity light field.  The
5$\lambda$/2 (\SI{583}{\nm}) cavity is made of Al$_{0.3}$Ga$_{0.7}$As. Photons
in the cavity are strongly coupled to quantum-well excitons to form mixed
light-matter bosonic polaritons.

The quasi continuous wave pump is a single-mode Ti:Sapphire laser tuned to the
first Bragg mode \SI{\sim 100}{\milli\electronvolt} above the polariton
energy.  A spatial light modulator is used to spatially pattern the pump beam
into a square lattice. A 0.4 NA objective is used for imaging the pattern onto
the sample. A cooled CCD and a \SI{0.55}{\meter} spectrometer is used for
imaging and energy resolving the emission.

The nonresonant excitation patterned in a square
geometry~\cite{ohadi_spontaneous_2015} initially creates a plasma of hot
electrons which then relax in energy and form bound excitons. The excitons in
this `reservoir' eventually relax to form polaritons. Polaritons are repelled
from reservoirs generated at each pump spot due to repulsive exciton-polariton
and polariton-polariton interactions. The resulting repulsive potential causes
polaritons to roll off and gather at the minima of the square potential.
Polaritons condense in the ground state by stimulated scattering once the
density at each center surpasses a critical threshold and form a macroscopic
state in each optical
trap~\cite{tsotsis_lasing_2012,cristofolini_optical_2013,askitopoulos_polariton_2013}.

As the trapped condensate is moved across the semiconductor sample, the energy
of the condensate varies by a standard deviation of $\SI{\sim 30}{\micro\eV}$
(see Appendix~\ref{si:disordpot}). This  $\SI{\sim 30}{\micro\eV}$ disorder potential
is approximately 10\% of the confining potential from the optical
trap~\cite{askitopoulos_polariton_2013}, which remains relatively constant after
condensation.

\section{Narrowing of momentum}

\begin{figure}
		\centering
		\includegraphics[width=.8\linewidth]{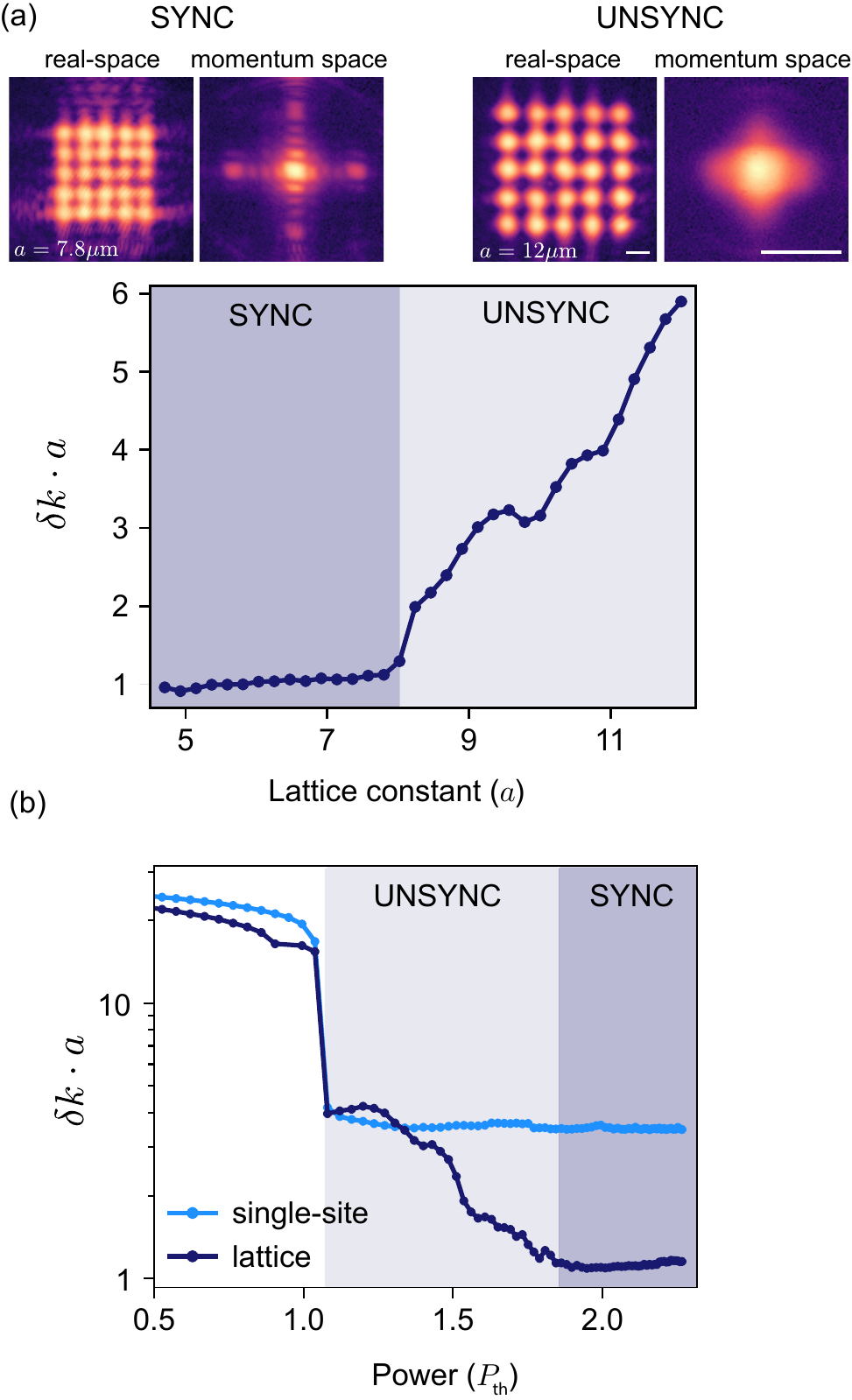}
		\caption{{Synchronization phase diagram.} Dependence of $\delta k \cdot a$
		versus \capind{a} lattice constant, and \capind{b} power. Top panel in \capind{a} shows
		real-space and momentum-space intensity (log scale) at $a=$\SI{7.8}{\um} (left) and $a=$\SI{11.8}{\um}. \capind{b} Power dependence comparison of $\delta k \cdot a$
		for full lattice (dark blue) and a spatially filtered single condensate (light blue). Real-space scale
		is \SI{10}{\um} and momentum space scale is
		\SI{1}{\per\um}.}
		\label{fig:2}
\end{figure}

Condensate lattices are created by patterning the optical excitation into a
square lattice of 6$\times$6 Gaussian spots (marked by grey circles in
Fig.~\ref{fig:1}c,d) using a spatial light modulator. The lattice constant ($a$)
of the trapping potential can be continuously varied while monitoring the total
emission of the condensate lattice in  momentum space (see also Supplemental
Movie 1). At low excitation powers incoherent polaritons confined by the
highest points of the lattice potential (dark spots in Fig.~\ref{fig:1}b) are
created with a very broad momentum distribution (full width half max = $\delta
k\simeq \SI{2.1}{\um^{-1}} $, $k$ is the in-plane polariton momentum). At a
threshold power ($P=\pth$) polaritons condense in the lattice sites. When the
condensates are far apart ($a>\SI{12}{\um}$) and the coupling between them is
small we observe a broad peak ($\delta k\simeq \SI{0.5}{\um^{-1}} $) centered at
zero momentum with no structure, and of a similar width to the emission from a
single condensate (Fig~\ref{fig:1}c). However, this changes as the lattice
constant is reduced to $a\simeq\SI{8}{\um}$ when a new diffraction pattern is
observed. It has a narrow center peak ($\delta k\simeq \SI{0.14}{\um^{-1}} $),
with 4 primary interference maxima at opposite corners of the reciprocal lattice
with $\mathbf{k}\cdot \mathbf{a}= 2\pi \times 0.96 $, suggesting the lattice is
phase locked (Fig.~\ref{fig:1}d, and also Appendix~\ref{si:phen}).

Tracking the width of the zero-momentum peak ($\delta k$) with lattice
constant ($a$) reveals the synchronization crossover (Fig.~\ref{fig:2}). For
$a>\SI{12}{\um}$, the condensates are weakly coupled and $\delta k$ is fixed to
that of a single lattice site. For $8<a<\SI{12}{\um}$,  $\delta k$ reduces as
the lattice constant $a$ decreases until reaching a value of $\delta
k \cdot a\sim 1$ at $a= \SI{8}{\um}$. This is the value expected for a 5-slit
diffraction grating, indicating phase coherence across the entire 5$\times$5
lattice. For smaller separations, $\delta k \cdot a$ remains constant until $a
\simeq\SI{4}{\um}$ where the trap size becomes comparable to the reservoir
width. In this regime, the traps merge into a single large inhomogeneous pump
spot.

The power dependence of $\delta k$ for a fixed lattice constant ($a=\SI{8}{\um}$)
shows the evolution of the long-range coherence with increasing density
(Fig.~\ref{fig:2}b, also Supplemental Movie 2). We compare this to emission from a single condensate
using spatial filtering for the same conditions. As the lattice power approaches
the single-site condensation threshold ($P=\pth$) polaritons condense in each
lattice site and $\delta k$ drops sharply. At threshold there is a condensate at
each site and the zero-momentum peak has a width similar to that of the spatially
filtered single condensate, $\delta k_1$. This is because there is no long-range
phase correlation and the lattice is in the unsynchronized phase. However, as power
increases further, $\delta k$ gradually decreases until it plateaus to $\delta k
\simeq \delta k_1/4$, at $P \simeq 2 \pth$. This value again matches with a 5-slit
grating (see SI.\ref{si:phen}), indicating long-range phase correlation of a
delocalized fluid across the entire lattice.

\begin{figure*}
		\centering
		\includegraphics[width=1\linewidth]{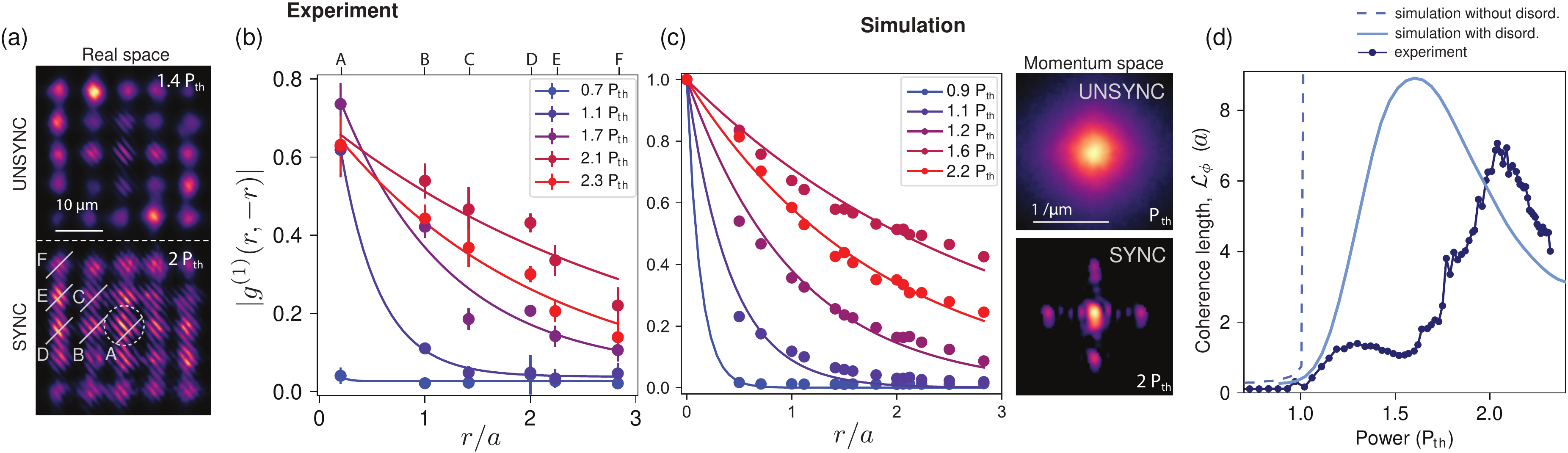}
		\caption{{First order correlation function.} \capind{a}
		Real-space interferograms for the
		unsynchronized (top) and synchronized  (bottom) phases. \capind{b,c} Power dependence of
		$\vert g^{(1)}(r,-r) \vert$ at various lattice position $r$, as a function of total
		lattice power in \capind{b} experiment and \capind{c} simulations. In the experiment, an average is taken over six site separations marked
		A-F and their corresponding mirrors in \capind{a}, with separations of
		$(0.5,2,\sqrt{2},4,2\sqrt{5},4\sqrt{2})a$. In simulations, the average is extracted for all site separations in the lattice. Lines are exponential fits. Panels to the right of \capind{c} are the time-averaged momentum space emission for UNSYNC ($P=\pth$) and SYNC ($P=2\pth$) phases. The simulated disorder ratio is 10\%. \capind{d} Coherence length ($\mathcal{L}_\phi$) vs power in simulations (with and without 10\% disorder) and in experiment. Simulation curves are averages of 25 randomly generated disorder potentials.}
\label{fig:3}
\end{figure*}

\section{Buildup of long-range coherence and collapse of energy detuning}

\begin{figure}
		\centering
		\includegraphics[width=1\linewidth]{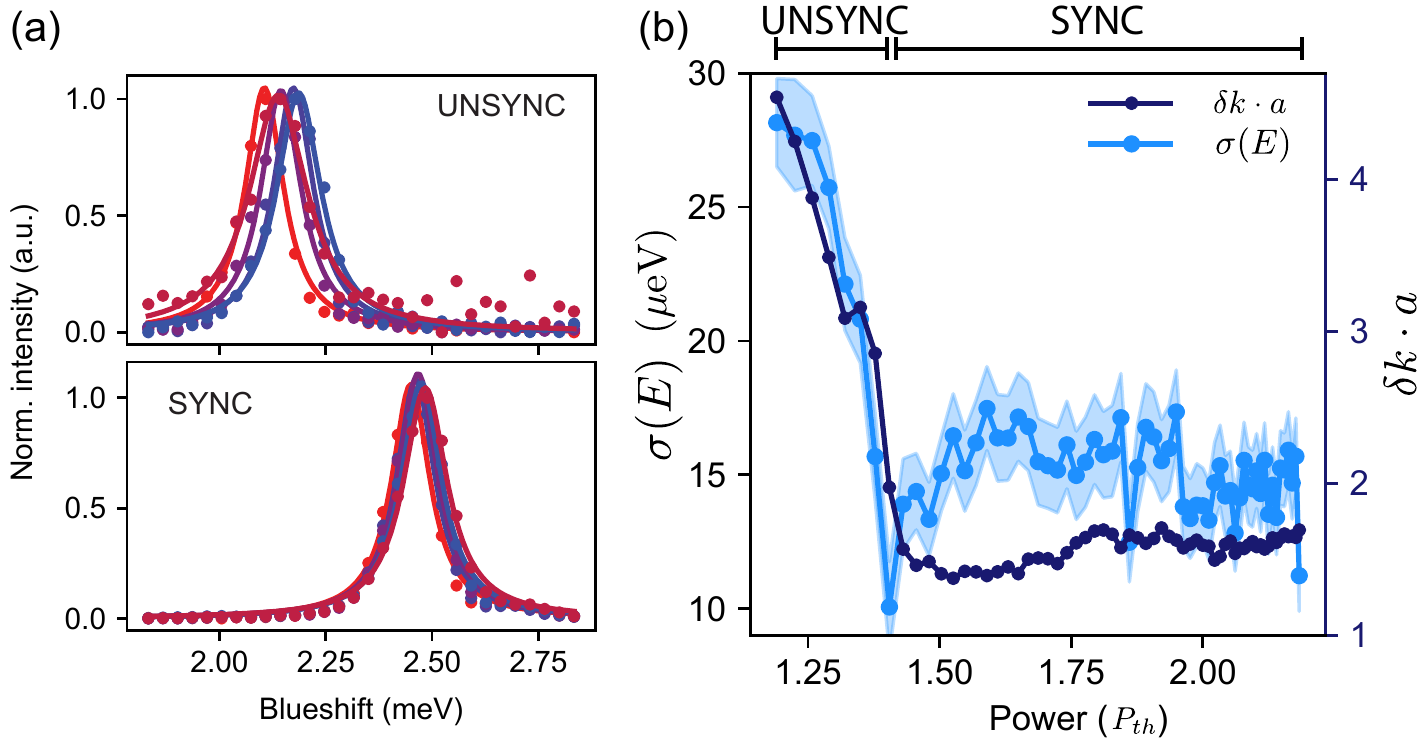}
		\caption{{Condensate energies as a function of power.} \capind{a} Energy
		spectra of a column of 5 condensates crossing the center of the lattice
        for $P=\pth$ (top panel) and $P=2\pth$ (bottom
		panel). \capind{b} Standard deviation of condensate energies ($\sigma(E)$) as a
		function of power (light blue) and the corresponding $\delta k \cdot a$
		(dark blue).
}\label{fig:sddev}
\end{figure}

The gradual decrease in momentum width $\delta k$ with power implies that phase
correlations in the lattice continuously increase with power as the coherence
length of the order parameter $\psi$ gradually increases. It has been shown that
in a driven-dissipative condensate with disorder, the phase correlation length
$\mathcal{L}_\phi$, over which the first order correlation function drops by $1/e$,
increases with Josephson coupling and condensates
density~\cite{janot_superfluid_2013}. To confirm this we measure the spatial
first-order coherence function $g^{(1)}$ at zero time delay. The lattice
emission is sent to a modified Mach-Zehnder interferometer with a retroreflector
in one arm. This interferes emission from opposite points relative to the
central condensate (placing the origin at the white dashed circle in
Fig.~\ref{fig:3}a) so that emission from $r$ interferes with $-r$. The contrast
of these interferograms then gives $\vert g^{(1)}(r,-r) \vert = \vert
g^{(1)}(-r,r) \vert$ with $\mathcal{L}_\phi$ being the phase correlation length
where $g^{(1)}$ drops to $1/e$. The interferograms (Fig.~\ref{fig:3}a) clearly
differ between the localized (top panel, $P=1.4 \pth$) and the delocalized phase
(bottom panel, $P=2 \pth$). Single-site coherence builds up immediately after
condensation. However, the coherence between different lattice sites turns on at
a power that increases with separation (see also Supplemental Movie 3). This first-order spatial coherence
function $g^{(1)}(-r,r)$ clearly changes at different powers
(Fig.~\ref{fig:3}b). Before condensation ($P<\pth$), $\mathcal{L}_\phi\simeq0$.
At single-site condensation (UNSYNC, $P=1.1 \pth$), order is confined to one site.
As power increases ($1.1 \pth<P<2.1 \pth$), $\mathcal{L}_\phi$ increases
gradually until it reaches $\sim$8 lattice constants (at $P=2.1 \pth$), but decays
at higher powers due to the appearance of higher order modes
(Fig.~\ref{fig:3}d). We note that the increase of $g^{(1)}$ (and the
reduction of $\delta k$) in steps is likely due to a percolation effect
where domains of condensates phase-lock in discrete steps.

Measuring the site energies of condensates in the synchronization
crossover reveals the reason behind this behavior. At threshold, condensates
form at slightly different energies (Fig.~\ref{fig:sddev}a). The energy
distribution observed in the UNSYNC phase is a result of spatial
inhomogeneities in the sample. However, as the power increases the condensate
energies converge as long-range order propagates across the lattice, and at
the SYNC phase they condense to one energy (Fig.~\ref{fig:sddev}b).

\section{Theoretical model and numerical simulations}
	\label{si:model}

\begin{figure*}
		\centering
		\includegraphics[width=1\linewidth]{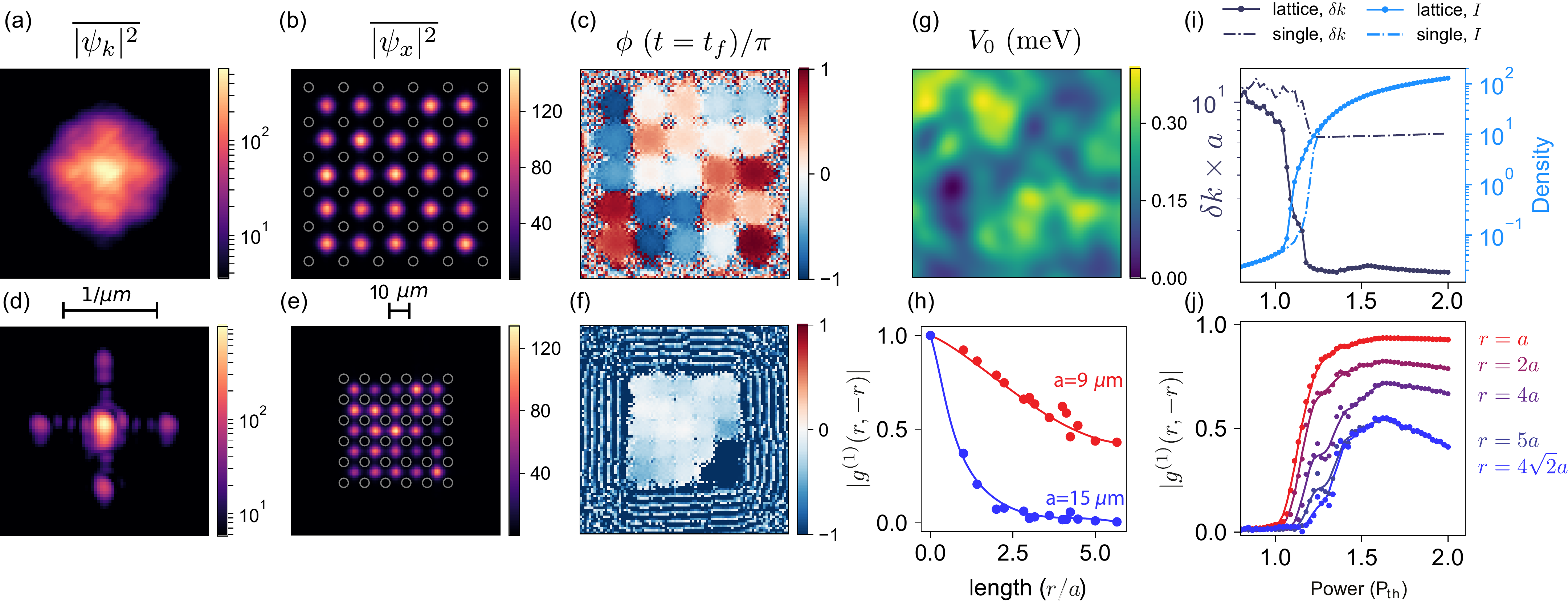}
		\caption{Time-averaged momentum space, real space and phase of an unsynchronized
		lattice (a-c) and synchronized lattice (d-f). (g) Static potential accounting for
		sample inhomogeneity. (h) $g^{(1)}$ versus length for UNSYNC (blue) and
		SYNC phase (red). (i) Power dependence of density (light blue)
		and momentum width (dark blue) for a synchronized lattice (solid line)
		and a single site (dotted line). (j) Power dependence of $g^{(1)}$ for
		different lattice lengths.
}\label{fig:5x5sim}
\end{figure*}

Our system can be modelled using the stochastic
driven-dissipative Gross-Pitaevskii (GP) equation which is a nonlinear Schr\"odinger equation
including pumping, dissipation, energy relaxation and Langevin noise to describe
fluctuations:
\begin{align}
    i \hbar \frac{d\psi}{dt}=&\left(-\frac{\hbar^2 \nabla^2}{2 m}+V_0+g_R n +g_p P+\alpha \vert \psi \vert^2\right)\psi,\nonumber\\
    &+\frac{i}{2}(\hbar R n-\hbar \gamma)\psi-i\hbar\Lambda \psi+\hbar \sqrt{\gamma+R n}\frac{dW}{dt}\\
    \frac{dn}{dt}=&P-(\gamma_R + R\vert \psi \vert^2) n.
    \label{eq:GP}
\end{align}
Here, $\psi$ is the wavefunction, $n$ is the reservoir density, $m$ is the polariton mass, $R$ is the scattering rate from the reservoir to the condensate, $\gamma$ is the polariton decay rate, $g_R$ and $g_p$
describe energy repulsion due to the presence of the reservoir and pump,
$\alpha$ is the strength of polariton-polariton interactions, $P$ is the non-
resonant pumping rate, $\gamma_R$ is the reservoir decay rate and $\Lambda$
is a spatial dependent function matching the pump profile, which accounts for a
phenomenological energy relaxation. Langevin noise is given by $dW$ which is a
complex Gaussian random variable characterized by correlation function $\langle
dW^* dW\rangle=dt$. Here, we also add a static disorder potential $V_0$
accounting for the sample inhomogeneity. Disorder ratio is defined as
$DOR=\overline{V_0}/(g_Rn +g_P P+\alpha\vert\psi\vert^2).$ The first order correlation
function is given by:
\begin{equation}
g^{(1)}(\mathbf{r},\mathbf{-r})=\frac{\langle
\psi(\mathbf{r})^* \psi(\mathbf{-r}) \rangle}{\sqrt{\langle \vert
\psi(\mathbf{r}) \vert^2 \rangle \langle \vert \psi(\mathbf{-r}) \vert^2
\rangle}}.
\end{equation}

Simulating the condensate lattice in a randomly generated disorder potential
demonstrates the crucial role of disorder in the synchronization crossover, as
shown in Fig.~\ref{fig:5x5sim}.
\bibnote[Parameters]{$\alpha=\SI{3}{\ueV\um^2}$; $g_r=2\alpha$; $g_P=0.6\alpha$; $\Lambda=0.2$;
$m^*=5.1\times10^{-5}m_e$; $R=\SI{0.01}{\ps^{-1}\um^2}$;
$\gamma=\SI{0.1}{\ps^{-1}\um^2}$; $\gamma_R=62.5\gamma$;}
Here, the disorder ratio ($DOR$) defined as the
mean ratio of a Gaussian-distributed disorder potential (Fig.~\ref{fig:5x5sim}g)
to the confinement potential by the pump pattern is ${\sim}10\%$ (as in the
experiment). The condensate lattice crosses from unsynchronized to synchronized
as the lattice constant is shrunk, with similar signatures to the experiment. In
the UNSYNC phase when $a=\SI{15}{\um}$ [Fig.~\ref{fig:5x5sim}(a-c)], momentum-space
peak is broad, condensates phases are unlocked and $\mathcal{L}_\phi\simeq a$
(blue line in Fig.~\ref{fig:5x5sim}h). This changes in the SYNC phase as the lattice
is shrunk to $a=\SI{9}{\um}$: the momentum-space peak narrows and new
diffraction peaks appear, condensates phase-lock [Fig.~\ref{fig:5x5sim}(d-f)],
and $\mathcal{L}_\phi$ goes beyond the lattice dimensions (red line in
Fig.~\ref{fig:5x5sim}h). The power dependence of the SYNC lattice shows gradual
narrowing of the momentum peak and buildup of long-range coherence
[Fig.~\ref{fig:5x5sim}(i,j)].

We find that the power dependence of correlation function strongly depends on
the spatial profile of the disorder potential. For this reason,  $\vert g^{(1)}
\vert$ simulations are averaged for 25 randomly-generated disorder potentials,
as shown in Fig.~\ref{fig:3}c. The general trend stays the same: in the UNSYNC
phase $g^{(1)}$ does not expand beyond a single site, whereas in the SYNC phase
$g^{(1)}$ builds up throughout the lattice with an exponential drop off
(Fig.~\ref{fig:3}c,d) (see Appendix~\ref{si:largelattice} for larger lattices).
In the synchronized phase, the Josephson coupling is strong enough to overcome
the disorder potential and phase-lock the condensates to form a single
macroscopic state. Without static disorder this synchronization crossover is
sharp and the coherence length is 3 orders of magnitude longer than in the
experiment (Fig.~\ref{fig:3}d). The minimization of the
width of this synchronization crossover with power can thus be used to search
for an optical potential that minimizes the static disorder potential, which has
applications in enabling using such arrays as simulators. We note that disorder
is inherent to any supported array of condensates, suggesting these observations
will be universal to all implementations.


\section{Concluding remarks}

We demonstrated a controllable crossover from unsynchronized to synchronized in
a driven-dissipative polariton condensate lattice with a background disorder
potential. The crossover occurs either by increasing the density of all
condensates or the nearest-neighbor Josephson coupling. The synchronized
(delocalized) phase is accompanied by the appearance of long-range order and
narrowing of the central peak in momentum space, whereas in the unsynchronized
phase order remains local, and the momentum space resembles the emission from a
single condensate. Using simulations of driven-dissipative GP equations, we find
that in the absence of disorder a sharp synchronization transition with
long-range spatial coherence is observed. However, the introduction of disorder
results in a softer crossover from unsynchronized to synchronized regimes with
finite range for the spatial coherence.

In thermalized cold atoms in a spatially disordered Bose-Hubbard potential,
there is the compressible Bose-glass (insulator) crossover phase between the
Mott insulator and superfluid phases~\cite{fisher_boson_1989,
krauth_superfluid-insulator_1991, fallani_ultracold_2007, meldgin_probing_2016}.
There, although the condensates share a global chemical potential there is no
long-range spatial coherence. Instead, the system forms puddles or domains of
coherent condensates, but there is no coherence between neighboring puddles. As
the tunneling rate exceeds the disorder potential, the superfluid puddles
coalesce until there exists a global superfluid with a spatial coherence length
that exceeds the size of the system.

By contrast, in the nonequilibrium case studied here, in the crossover regime
the puddles or domains have different chemical potentials, or emission energies.
Furthermore, in the superfluid phase the spatial coherence has a finite range
that increases with pump power and tunneling rate. Hence driven-dissipative
generalization of superfluidity can only be observed in a system of finite size,
as discussed theoretically in ref.~\cite{janot_superfluid_2013}. Unless actively
compensated~\cite{ohadi_spin_2017}, this may have consequences for the scale-up
of simulators based on lattices of exciton-polariton condensates.

With our capability to optically compensate for disorder~\cite{ohadi_spin_2017},
it would be interesting to study the interplay between on-demand disorder
strengths and the synchronization crossover, and their relation to the
characteristic length-scales of the first-order spatial correlation. In addition
to fundamental interest, synchronization of arrays of polariton lasers for
example, would be advantageous for creating high-power density coherent sources
at low pump density. Conventional laser diodes locked by injection coupling can
synchronize only in a narrow range of parameters~\cite{winful_stability_1988},
due to their large carrier induced red-shift. By contrast, we demonstrate that
with a carrier induced blue-shift, synchronization readily occurs, suggesting
that polariton laser approaches may be highly profitable.


\section*{Acknowledgments} We thank Jonathan Keeling, Peter Kirton and Ulrich
Schneider for fruitful discussions. We acknowledge Grants No. EPSRC
EP/L027151/1, ERC LINASS 320503, and Leverhulme Trust Grant No. VP1-2013-011. PS
acknowledges support from ITMO Fellowship Program and megaGrant No.
14.Y26.31.0015 of the Ministry of Education and Science of Russian Federation
and Greece-Russia bilateral `POLISIMULATOR' project on Quantum Technologies
funded by Greek GSRT. PE acknowledges support from SFI Grant No. 15/IACA/3402.
TL was supported by the Singaporean Ministry of Education (MOE2017-T2-1-001).


%
%

\appendix
\section{Disorder potential}
\label{si:disordpot}
\begin{figure}
		\centering
		\includegraphics[width=.7\linewidth]{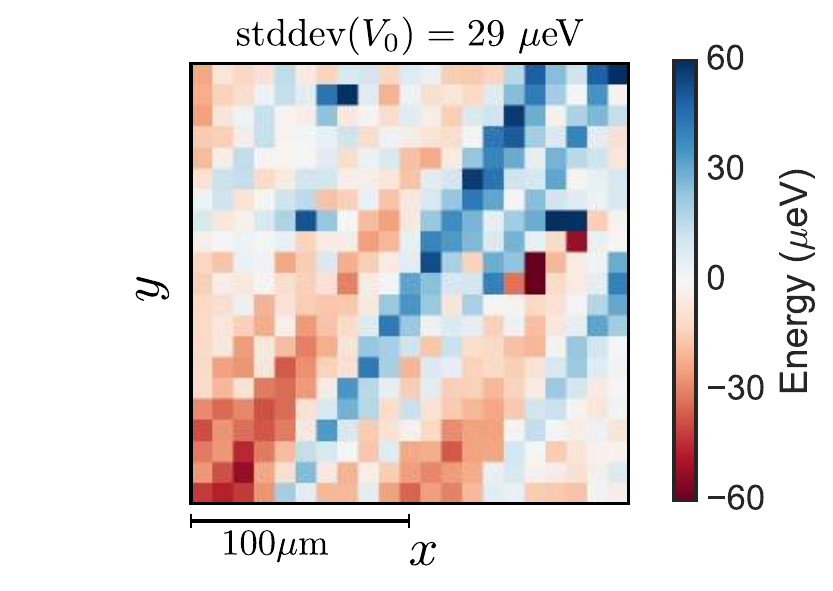}
		\caption{Background disorder potential}\label{fig:energymap}
\end{figure}

Background disorder potential $V_0$ is measured experimentally by moving a
single trapped condensate over the sample and measuring its energy, as shown in
Fig.~\ref{fig:energymap}. Standard deviation of disorder potential is $\SI{\sim
29}{\ueV}$.

\section{Phenomenological model}
\label{si:phen}

\begin{figure}[h]
	\centering
    \includegraphics[width=1\linewidth]{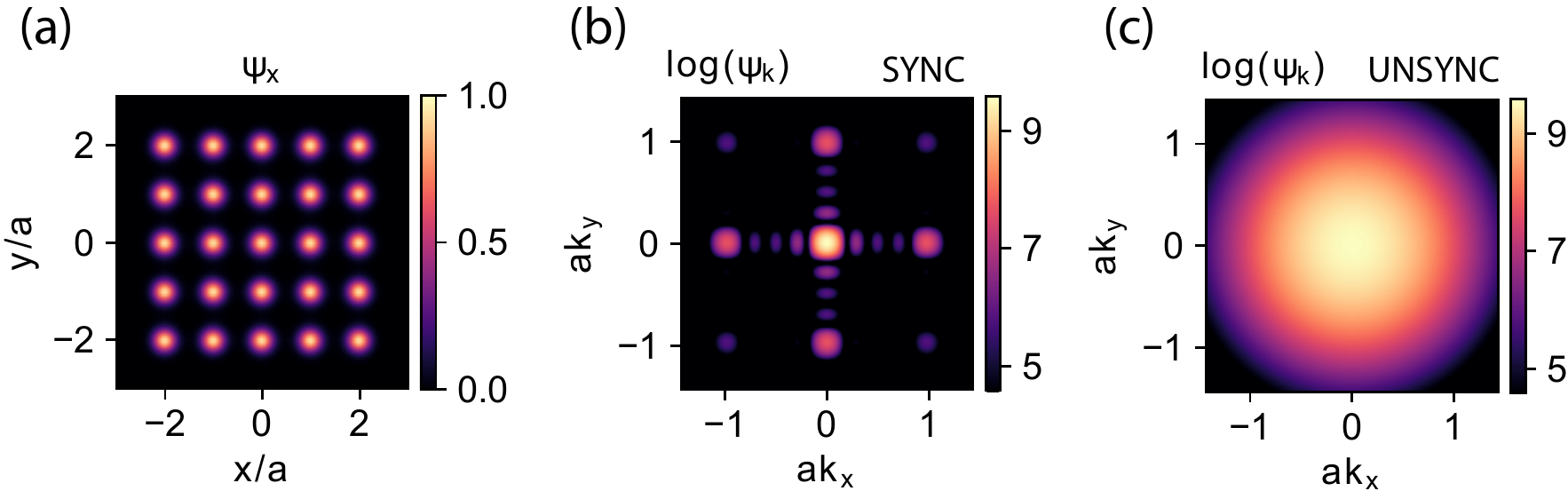}
    \caption{Intensity of the synchronized phase in real (a) and reciprocal (b)
    space. (c) Intensity of the UNSYNC phase in momentum space. The only parameter
    is $L/a = 0.22$.}
    \label{fig:phen}
\end{figure}

Let us assume a coherent superposition of Gaussian wavefunctions localized at
each lattice point:
\begin{equation}
    A \sum_{n,m} e^{-\left((x-na)^2+(y-ma)^2\right)/(2L^2)},
\end{equation}
where $A$ is an overall amplitude, $a$ is the lattice constant, and $L$ defines
the width of each Gaussian. In reciprocal space the wavefunction is given by the
Fourier transform:
\begin{align}
\tilde{\psi}(k_x,k_y)&=\frac{1}{2\pi}\int^\infty_{-\infty}\int^\infty_{-\infty}\psi(x,y)e^{i(k_x x + k_y y)}dxdy\nonumber\\
&=AL^2 e^{-(k_x^2+k_y^2)L^2/2}\sum_{n,m}e^{i a(k_xn+k_ym)}.
\label{eq:wfsf}
\end{align}
The corresponding intensities of the wavefunctions in real and reciprocal space
are shown in Fig.~\ref{fig:phen}. In the UNSYNC phase there is no coherence between
different Gaussian spots. In this case the intensity in real space is given by:
\begin{equation}
    I(x,y)=\vert A \vert^2 \sum_{n,m} e^{-\left((x-na)^2+(y-ma)^2\right)/L^2}.
\end{equation}
The intensity in reciprocal space is obtained taking the Fourier transform of
each Gaussian spot separately, and summing the intensities rather than the
amplitudes:
\begin{equation}
    \tilde{I}(k_x,k_y)=NM\vert A \vert^2 L^4 e^{-(k_x^2+k_y^2)/L^2}.
\label{eq:wfin}
\end{equation}
Since there is no coherence between Gaussian spots, the interference term
appearing in Eq.~\ref{eq:wfsf} is no longer present in Eq.~\ref{eq:wfin}. The
sum over $n$ and $m$ has been reduced to the total number of spots, $NM$, in the
square lattice. The corresponding intensity of the wavefunctions in the momentum
space is shown in Fig.~\ref{fig:phen}c. This model obviously cannot describe
why the crossover is not sharp and why the coherence length decays over length.
For that we model the system using a stochastic driven-dissipative
GP equation.

\begin{figure}
	\centering
	\includegraphics[width=0.7\linewidth]{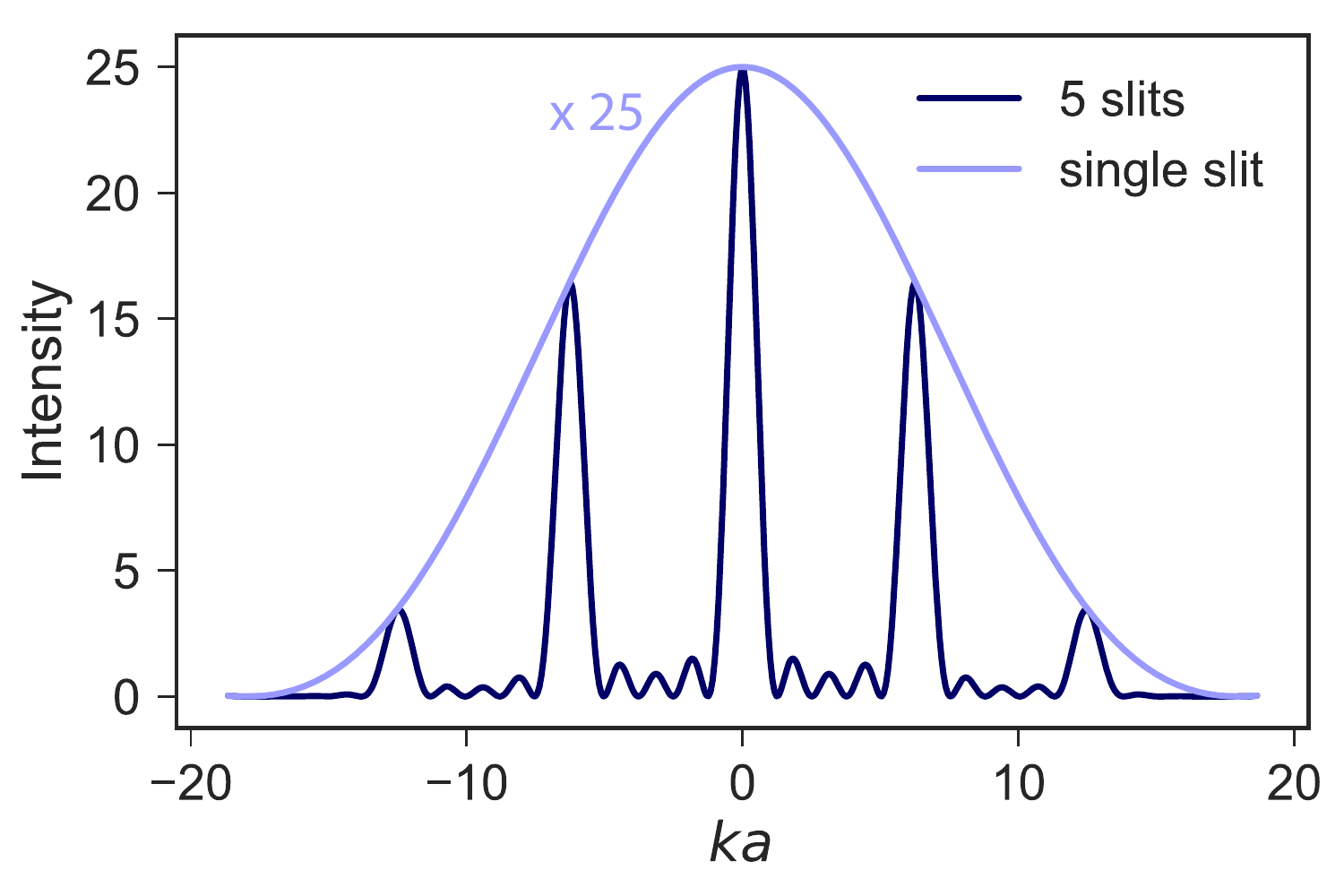}
	\caption{Fraunhofer diffraction of a single slit (light blue) and 5 slits
	(dark blue) with $b/a=0.35$.} \label{sig:si-Nslit}
\end{figure}

\begin{figure}
		\centering
	    \includegraphics[width=.9\linewidth]{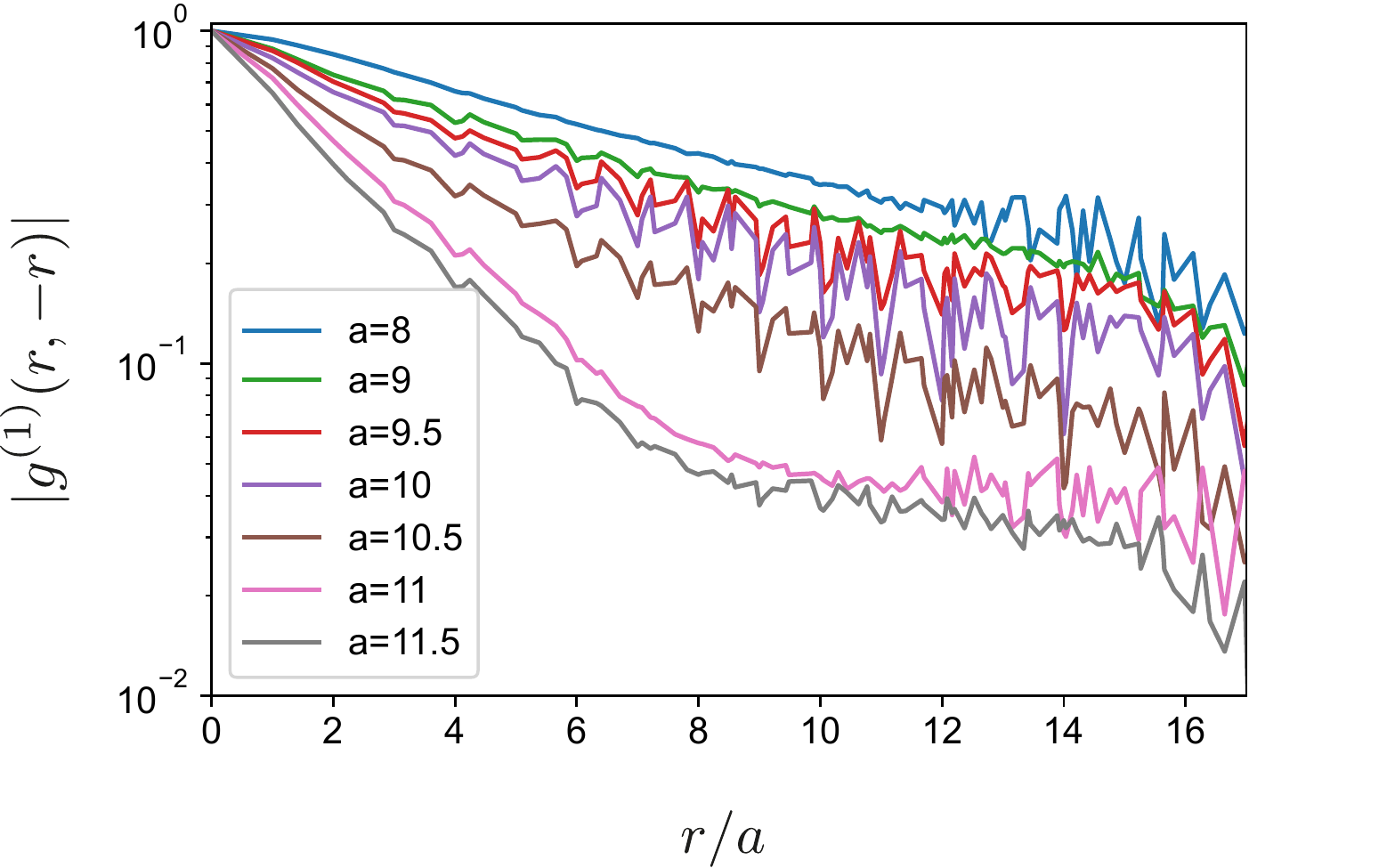}
	    \caption{Average first order correlation function $g^{(1)}(r/a)$ vs.
	    length $r$ for various lattice constants of a 15$\times$15 condensate
	    lattice.}
	    \label{fig:g1-large-lattice}
\end{figure}

We note here the similarities between the synchronized lattice intensity profile, and
Fraunhofer diffraction from many slits given by~\cite{born_principles_1999}:
\begin{equation}
I=I_0\left(\frac{\sin{\beta}}{\beta}\right)^2 \left(\frac{\sin{N\alpha}}{\sin{\alpha}}\right)^2,
\label{eq:si-slitdiff}
\end{equation}
where $\beta=\frac{1}{2} k_\parallel b$, $\alpha=\frac{1}{2} k_\parallel a$.
Here, $a$ is the slit separation, $b$ is the slit width, $N$ is the number of
slits and $I_0$ includes all the constants to give maximum intensity from a
single slit (Fig.~\ref{sig:si-Nslit}). From Eq.~\ref{eq:si-slitdiff}, we can see that
the first intensity minimum occurs at $ka=2\pi/N$. For 5 slits, full width at half
maximum of center peak is $\delta k \cdot a\simeq1.13$.

\section{Large lattice simulations with disorder} 
\label{si:largelattice}

The dependence of first order correlation function with the lattice constant for
a $15\times15$ lattice in a random disorder potential ($DOR=10\%$) shows an
exponential decay with a coherence length $\mathcal{L}_\phi$ which grows as the
lattice constant $a$ reduces, as shown in Fig.~\ref{fig:g1-large-lattice}.

\bibliography{bib}

\begin{thebibliography}{35}%
\makeatletter
\providecommand \@ifxundefined [1]{%
 \@ifx{#1\undefined}
}%
\providecommand \@ifnum [1]{%
 \ifnum #1\expandafter \@firstoftwo
 \else \expandafter \@secondoftwo
 \fi
}%
\providecommand \@ifx [1]{%
 \ifx #1\expandafter \@firstoftwo
 \else \expandafter \@secondoftwo
 \fi
}%
\providecommand \natexlab [1]{#1}%
\providecommand \enquote  [1]{``#1''}%
\providecommand \bibnamefont  [1]{#1}%
\providecommand \bibfnamefont [1]{#1}%
\providecommand \citenamefont [1]{#1}%
\providecommand \href@noop [0]{\@secondoftwo}%
\providecommand \href [0]{\begingroup \@sanitize@url \@href}%
\providecommand \@href[1]{\@@startlink{#1}\@@href}%
\providecommand \@@href[1]{\endgroup#1\@@endlink}%
\providecommand \@sanitize@url [0]{\catcode `\\12\catcode `\$12\catcode
  `\&12\catcode `\#12\catcode `\^12\catcode `\_12\catcode `\%12\relax}%
\providecommand \@@startlink[1]{}%
\providecommand \@@endlink[0]{}%
\providecommand \url  [0]{\begingroup\@sanitize@url \@url }%
\providecommand \@url [1]{\endgroup\@href {#1}{\urlprefix }}%
\providecommand \urlprefix  [0]{URL }%
\providecommand \Eprint [0]{\href }%
\providecommand \doibase [0]{http://dx.doi.org/}%
\providecommand \selectlanguage [0]{\@gobble}%
\providecommand \bibinfo  [0]{\@secondoftwo}%
\providecommand \bibfield  [0]{\@secondoftwo}%
\providecommand \translation [1]{[#1]}%
\providecommand \BibitemOpen [0]{}%
\providecommand \bibitemStop [0]{}%
\providecommand \bibitemNoStop [0]{.\EOS\space}%
\providecommand \EOS [0]{\spacefactor3000\relax}%
\providecommand \BibitemShut  [1]{\csname bibitem#1\endcsname}%
\let\auto@bib@innerbib\@empty
\bibitem [{\citenamefont {Kavokin}\ \emph {et~al.}(2007)\citenamefont
  {Kavokin}, \citenamefont {Baumberg}, \citenamefont {Malpuech},\ and\
  \citenamefont {Laussy}}]{kavokin_microcavities_2007}%
  \BibitemOpen
  \bibfield  {author} {\bibinfo {author} {\bibfnamefont {Alexey~V.}\
  \bibnamefont {Kavokin}}, \bibinfo {author} {\bibfnamefont {Jeremy}\
  \bibnamefont {Baumberg}}, \bibinfo {author} {\bibfnamefont {Guillaume}\
  \bibnamefont {Malpuech}}, \ and\ \bibinfo {author} {\bibfnamefont
  {Fabrice~P.}\ \bibnamefont {Laussy}},\ }\href@noop {} {\emph {\bibinfo
  {title} {Microcavities}}}\ (\bibinfo  {publisher} {Oxford Univ. Press},\
  \bibinfo {address} {Oxford},\ \bibinfo {year} {2007})\BibitemShut {NoStop}%
\bibitem [{\citenamefont {Carusotto}\ and\ \citenamefont
  {Ciuti}(2013)}]{carusotto_quantum_2013}%
  \BibitemOpen
  \bibfield  {author} {\bibinfo {author} {\bibfnamefont {Iacopo}\ \bibnamefont
  {Carusotto}}\ and\ \bibinfo {author} {\bibfnamefont {Cristiano}\ \bibnamefont
  {Ciuti}},\ }\bibfield  {title} {\enquote {\bibinfo {title} {Quantum fluids of
  light},}\ }\href {\doibase 10.1103/RevModPhys.85.299} {\bibfield  {journal}
  {\bibinfo  {journal} {Rev. Mod. Phys.}\ }\textbf {\bibinfo {volume} {85}},\
  \bibinfo {pages} {299--366} (\bibinfo {year} {2013})}\BibitemShut {NoStop}%
\bibitem [{\citenamefont {Lai}\ \emph {et~al.}(2007)\citenamefont {Lai},
  \citenamefont {Kim}, \citenamefont {Utsunomiya}, \citenamefont {Roumpos},
  \citenamefont {Deng}, \citenamefont {Fraser}, \citenamefont {Byrnes},
  \citenamefont {Recher}, \citenamefont {Kumada}, \citenamefont {Fujisawa},\
  and\ \citenamefont {Yamamoto}}]{lai_coherent_2007}%
  \BibitemOpen
  \bibfield  {author} {\bibinfo {author} {\bibfnamefont {C.~W.}\ \bibnamefont
  {Lai}}, \bibinfo {author} {\bibfnamefont {N.~Y.}\ \bibnamefont {Kim}},
  \bibinfo {author} {\bibfnamefont {S.}~\bibnamefont {Utsunomiya}}, \bibinfo
  {author} {\bibfnamefont {G.}~\bibnamefont {Roumpos}}, \bibinfo {author}
  {\bibfnamefont {H.}~\bibnamefont {Deng}}, \bibinfo {author} {\bibfnamefont
  {M.~D.}\ \bibnamefont {Fraser}}, \bibinfo {author} {\bibfnamefont
  {T.}~\bibnamefont {Byrnes}}, \bibinfo {author} {\bibfnamefont
  {P.}~\bibnamefont {Recher}}, \bibinfo {author} {\bibfnamefont
  {N.}~\bibnamefont {Kumada}}, \bibinfo {author} {\bibfnamefont
  {T.}~\bibnamefont {Fujisawa}}, \ and\ \bibinfo {author} {\bibfnamefont
  {Y.}~\bibnamefont {Yamamoto}},\ }\bibfield  {title} {\enquote {\bibinfo
  {title} {Coherent zero-state and [pgr]-state in an exciton-polariton
  condensate array},}\ }\href {\doibase 10.1038/nature06334} {\bibfield
  {journal} {\bibinfo  {journal} {Nature}\ }\textbf {\bibinfo {volume} {450}},\
  \bibinfo {pages} {529--532} (\bibinfo {year} {2007})}\BibitemShut {NoStop}%
\bibitem [{\citenamefont {Galbiati}\ \emph {et~al.}(2012)\citenamefont
  {Galbiati}, \citenamefont {Ferrier}, \citenamefont {Solnyshkov},
  \citenamefont {Tanese}, \citenamefont {Wertz}, \citenamefont {Amo},
  \citenamefont {Abbarchi}, \citenamefont {Senellart}, \citenamefont {Sagnes},
  \citenamefont {Lemaître}, \citenamefont {Galopin}, \citenamefont
  {Malpuech},\ and\ \citenamefont {Bloch}}]{galbiati_polariton_2012}%
  \BibitemOpen
  \bibfield  {author} {\bibinfo {author} {\bibfnamefont {Marta}\ \bibnamefont
  {Galbiati}}, \bibinfo {author} {\bibfnamefont {Lydie}\ \bibnamefont
  {Ferrier}}, \bibinfo {author} {\bibfnamefont {Dmitry~D.}\ \bibnamefont
  {Solnyshkov}}, \bibinfo {author} {\bibfnamefont {Dimitrii}\ \bibnamefont
  {Tanese}}, \bibinfo {author} {\bibfnamefont {Esther}\ \bibnamefont {Wertz}},
  \bibinfo {author} {\bibfnamefont {Alberto}\ \bibnamefont {Amo}}, \bibinfo
  {author} {\bibfnamefont {Marco}\ \bibnamefont {Abbarchi}}, \bibinfo {author}
  {\bibfnamefont {Pascale}\ \bibnamefont {Senellart}}, \bibinfo {author}
  {\bibfnamefont {Isabelle}\ \bibnamefont {Sagnes}}, \bibinfo {author}
  {\bibfnamefont {Aristide}\ \bibnamefont {Lemaître}}, \bibinfo {author}
  {\bibfnamefont {Elisabeth}\ \bibnamefont {Galopin}}, \bibinfo {author}
  {\bibfnamefont {Guillaume}\ \bibnamefont {Malpuech}}, \ and\ \bibinfo
  {author} {\bibfnamefont {Jacqueline}\ \bibnamefont {Bloch}},\ }\bibfield
  {title} {\enquote {\bibinfo {title} {Polariton {Condensation} in {Photonic}
  {Molecules}},}\ }\href {\doibase 10.1103/PhysRevLett.108.126403} {\bibfield
  {journal} {\bibinfo  {journal} {Phys. Rev. Lett.}\ }\textbf {\bibinfo
  {volume} {108}},\ \bibinfo {pages} {126403} (\bibinfo {year}
  {2012})}\BibitemShut {NoStop}%
\bibitem [{\citenamefont {Wertz}\ \emph {et~al.}(2010)\citenamefont {Wertz},
  \citenamefont {Ferrier}, \citenamefont {Solnyshkov}, \citenamefont {Johne},
  \citenamefont {Sanvitto}, \citenamefont {Lemaître}, \citenamefont {Sagnes},
  \citenamefont {Grousson}, \citenamefont {Kavokin}, \citenamefont {Senellart},
  \citenamefont {Malpuech},\ and\ \citenamefont
  {Bloch}}]{wertz_spontaneous_2010}%
  \BibitemOpen
  \bibfield  {author} {\bibinfo {author} {\bibfnamefont {E.}~\bibnamefont
  {Wertz}}, \bibinfo {author} {\bibfnamefont {L.}~\bibnamefont {Ferrier}},
  \bibinfo {author} {\bibfnamefont {D.~D.}\ \bibnamefont {Solnyshkov}},
  \bibinfo {author} {\bibfnamefont {R.}~\bibnamefont {Johne}}, \bibinfo
  {author} {\bibfnamefont {D.}~\bibnamefont {Sanvitto}}, \bibinfo {author}
  {\bibfnamefont {A.}~\bibnamefont {Lemaître}}, \bibinfo {author}
  {\bibfnamefont {I.}~\bibnamefont {Sagnes}}, \bibinfo {author} {\bibfnamefont
  {R.}~\bibnamefont {Grousson}}, \bibinfo {author} {\bibfnamefont {A.~V.}\
  \bibnamefont {Kavokin}}, \bibinfo {author} {\bibfnamefont {P.}~\bibnamefont
  {Senellart}}, \bibinfo {author} {\bibfnamefont {G.}~\bibnamefont {Malpuech}},
  \ and\ \bibinfo {author} {\bibfnamefont {J.}~\bibnamefont {Bloch}},\
  }\bibfield  {title} {\enquote {\bibinfo {title} {Spontaneous formation and
  optical manipulation of extended polariton condensates},}\ }\href {\doibase
  10.1038/nphys1750} {\bibfield  {journal} {\bibinfo  {journal} {Nature Phys.}\
  }\textbf {\bibinfo {volume} {6}},\ \bibinfo {pages} {860--864} (\bibinfo
  {year} {2010})}\BibitemShut {NoStop}%
\bibitem [{\citenamefont {Takemura}\ \emph {et~al.}(2014)\citenamefont
  {Takemura}, \citenamefont {Trebaol}, \citenamefont {Wouters}, \citenamefont
  {Portella-Oberli},\ and\ \citenamefont
  {Deveaud}}]{takemura_polaritonic_2014}%
  \BibitemOpen
  \bibfield  {author} {\bibinfo {author} {\bibfnamefont {N.}~\bibnamefont
  {Takemura}}, \bibinfo {author} {\bibfnamefont {S.}~\bibnamefont {Trebaol}},
  \bibinfo {author} {\bibfnamefont {M.}~\bibnamefont {Wouters}}, \bibinfo
  {author} {\bibfnamefont {M.~T.}\ \bibnamefont {Portella-Oberli}}, \ and\
  \bibinfo {author} {\bibfnamefont {B.}~\bibnamefont {Deveaud}},\ }\bibfield
  {title} {\enquote {\bibinfo {title} {Polaritonic feshbach resonance},}\
  }\href {\doibase 10.1038/nphys2999} {\bibfield  {journal} {\bibinfo
  {journal} {Nature Phys.}\ }\textbf {\bibinfo {volume} {10}},\ \bibinfo
  {pages} {500--504} (\bibinfo {year} {2014})}\BibitemShut {NoStop}%
\bibitem [{\citenamefont {Gao}\ \emph {et~al.}(2012)\citenamefont {Gao},
  \citenamefont {Eldridge}, \citenamefont {Liew}, \citenamefont {Tsintzos},
  \citenamefont {Stavrinidis}, \citenamefont {Deligeorgis}, \citenamefont
  {Hatzopoulos},\ and\ \citenamefont {Savvidis}}]{gao_polariton_2012}%
  \BibitemOpen
  \bibfield  {author} {\bibinfo {author} {\bibfnamefont {T.}~\bibnamefont
  {Gao}}, \bibinfo {author} {\bibfnamefont {P.~S.}\ \bibnamefont {Eldridge}},
  \bibinfo {author} {\bibfnamefont {T.~C.~H.}\ \bibnamefont {Liew}}, \bibinfo
  {author} {\bibfnamefont {S.~I.}\ \bibnamefont {Tsintzos}}, \bibinfo {author}
  {\bibfnamefont {G.}~\bibnamefont {Stavrinidis}}, \bibinfo {author}
  {\bibfnamefont {G.}~\bibnamefont {Deligeorgis}}, \bibinfo {author}
  {\bibfnamefont {Z.}~\bibnamefont {Hatzopoulos}}, \ and\ \bibinfo {author}
  {\bibfnamefont {P.~G.}\ \bibnamefont {Savvidis}},\ }\bibfield  {title}
  {\enquote {\bibinfo {title} {Polariton condensate transistor switch},}\
  }\href {\doibase 10.1103/PhysRevB.85.235102} {\bibfield  {journal} {\bibinfo
  {journal} {Phys. Rev. B}\ }\textbf {\bibinfo {volume} {85}},\ \bibinfo
  {pages} {235102} (\bibinfo {year} {2012})}\BibitemShut {NoStop}%
\bibitem [{\citenamefont {Deng}\ \emph {et~al.}(2002)\citenamefont {Deng},
  \citenamefont {Weihs}, \citenamefont {Santori}, \citenamefont {Bloch},\ and\
  \citenamefont {Yamamoto}}]{deng_condensation_2002}%
  \BibitemOpen
  \bibfield  {author} {\bibinfo {author} {\bibfnamefont {Hui}\ \bibnamefont
  {Deng}}, \bibinfo {author} {\bibfnamefont {Gregor}\ \bibnamefont {Weihs}},
  \bibinfo {author} {\bibfnamefont {Charles}\ \bibnamefont {Santori}}, \bibinfo
  {author} {\bibfnamefont {Jacqueline}\ \bibnamefont {Bloch}}, \ and\ \bibinfo
  {author} {\bibfnamefont {Yoshihisa}\ \bibnamefont {Yamamoto}},\ }\bibfield
  {title} {\enquote {\bibinfo {title} {Condensation of semiconductor
  microcavity exciton polaritons},}\ }\href {\doibase 10.1126/science.1074464}
  {\bibfield  {journal} {\bibinfo  {journal} {Science}\ }\textbf {\bibinfo
  {volume} {298}},\ \bibinfo {pages} {199--202} (\bibinfo {year}
  {2002})}\BibitemShut {NoStop}%
\bibitem [{\citenamefont {Kasprzak}\ \emph {et~al.}(2006)\citenamefont
  {Kasprzak}, \citenamefont {Richard}, \citenamefont {Kundermann},
  \citenamefont {Baas}, \citenamefont {Jeambrun}, \citenamefont {Keeling},
  \citenamefont {Marchetti}, \citenamefont {Szymańska}, \citenamefont
  {André}, \citenamefont {Staehli} \emph
  {et~al.}}]{kasprzak_bose-einstein_2006}%
  \BibitemOpen
  \bibfield  {author} {\bibinfo {author} {\bibfnamefont {J.}~\bibnamefont
  {Kasprzak}}, \bibinfo {author} {\bibfnamefont {M.}~\bibnamefont {Richard}},
  \bibinfo {author} {\bibfnamefont {S.}~\bibnamefont {Kundermann}}, \bibinfo
  {author} {\bibfnamefont {A.}~\bibnamefont {Baas}}, \bibinfo {author}
  {\bibfnamefont {P.}~\bibnamefont {Jeambrun}}, \bibinfo {author}
  {\bibfnamefont {J~M~J}\ \bibnamefont {Keeling}}, \bibinfo {author}
  {\bibfnamefont {F~M}\ \bibnamefont {Marchetti}}, \bibinfo {author}
  {\bibfnamefont {M~H}\ \bibnamefont {Szymańska}}, \bibinfo {author}
  {\bibfnamefont {R.}~\bibnamefont {André}}, \bibinfo {author} {\bibfnamefont
  {J~L}\ \bibnamefont {Staehli}},  \emph {et~al.},\ }\bibfield  {title}
  {\enquote {\bibinfo {title} {Bose-einstein condensation of exciton
  polaritons.}}\ }\href {\doibase 10.1038/nature05131} {\bibfield  {journal}
  {\bibinfo  {journal} {Nature}\ }\textbf {\bibinfo {volume} {443}},\ \bibinfo
  {pages} {409--414} (\bibinfo {year} {2006})}\BibitemShut {NoStop}%
\bibitem [{\citenamefont {Baumberg}\ \emph {et~al.}(2008)\citenamefont
  {Baumberg}, \citenamefont {Kavokin}, \citenamefont {Christopoulos},
  \citenamefont {Grundy}, \citenamefont {Butté}, \citenamefont {Christmann},
  \citenamefont {Solnyshkov}, \citenamefont {Malpuech}, \citenamefont
  {Baldassarri Höger~von Högersthal}, \citenamefont {Feltin} \emph
  {et~al.}}]{baumberg_spontaneous_2008}%
  \BibitemOpen
  \bibfield  {author} {\bibinfo {author} {\bibfnamefont {J.}~\bibnamefont
  {Baumberg}}, \bibinfo {author} {\bibfnamefont {A.}~\bibnamefont {Kavokin}},
  \bibinfo {author} {\bibfnamefont {S.}~\bibnamefont {Christopoulos}}, \bibinfo
  {author} {\bibfnamefont {A.}~\bibnamefont {Grundy}}, \bibinfo {author}
  {\bibfnamefont {R.}~\bibnamefont {Butté}}, \bibinfo {author} {\bibfnamefont
  {G.}~\bibnamefont {Christmann}}, \bibinfo {author} {\bibfnamefont
  {D.}~\bibnamefont {Solnyshkov}}, \bibinfo {author} {\bibfnamefont
  {G.}~\bibnamefont {Malpuech}}, \bibinfo {author} {\bibfnamefont
  {G.}~\bibnamefont {Baldassarri Höger~von Högersthal}}, \bibinfo {author}
  {\bibfnamefont {E.}~\bibnamefont {Feltin}},  \emph {et~al.},\ }\bibfield
  {title} {\enquote {\bibinfo {title} {Spontaneous polarization buildup in a
  room-temperature polariton laser},}\ }\href {\doibase
  10.1103/PhysRevLett.101.136409} {\bibfield  {journal} {\bibinfo  {journal}
  {Phys. Rev. Lett.}\ }\textbf {\bibinfo {volume} {101}},\ \bibinfo {pages}
  {136409} (\bibinfo {year} {2008})}\BibitemShut {NoStop}%
\bibitem [{\citenamefont {Amo}\ \emph {et~al.}(2009)\citenamefont {Amo},
  \citenamefont {Lefrère}, \citenamefont {Pigeon}, \citenamefont {Adrados},
  \citenamefont {Ciuti}, \citenamefont {Carusotto}, \citenamefont {Houdré},
  \citenamefont {Giacobino},\ and\ \citenamefont
  {Bramati}}]{amo_superfluidity_2009}%
  \BibitemOpen
  \bibfield  {author} {\bibinfo {author} {\bibfnamefont {Alberto}\ \bibnamefont
  {Amo}}, \bibinfo {author} {\bibfnamefont {Jérôme}\ \bibnamefont
  {Lefrère}}, \bibinfo {author} {\bibfnamefont {Simon}\ \bibnamefont
  {Pigeon}}, \bibinfo {author} {\bibfnamefont {Claire}\ \bibnamefont
  {Adrados}}, \bibinfo {author} {\bibfnamefont {Cristiano}\ \bibnamefont
  {Ciuti}}, \bibinfo {author} {\bibfnamefont {Iacopo}\ \bibnamefont
  {Carusotto}}, \bibinfo {author} {\bibfnamefont {Romuald}\ \bibnamefont
  {Houdré}}, \bibinfo {author} {\bibfnamefont {Elisabeth}\ \bibnamefont
  {Giacobino}}, \ and\ \bibinfo {author} {\bibfnamefont {Alberto}\ \bibnamefont
  {Bramati}},\ }\bibfield  {title} {\enquote {\bibinfo {title} {Superfluidity
  of polaritons in semiconductor microcavities},}\ }\href {\doibase
  10.1038/nphys1364} {\bibfield  {journal} {\bibinfo  {journal} {Nature Phys.}\
  }\textbf {\bibinfo {volume} {5}},\ \bibinfo {pages} {805--810} (\bibinfo
  {year} {2009})}\BibitemShut {NoStop}%
\bibitem [{\citenamefont {Lerario}\ \emph {et~al.}(2017)\citenamefont
  {Lerario}, \citenamefont {Fieramosca}, \citenamefont {Barachati},
  \citenamefont {Ballarini}, \citenamefont {Daskalakis}, \citenamefont
  {Dominici}, \citenamefont {De~Giorgi}, \citenamefont {Maier}, \citenamefont
  {Gigli}, \citenamefont {Kéna-Cohen},\ and\ \citenamefont
  {Sanvitto}}]{lerario_room-temperature_2017}%
  \BibitemOpen
  \bibfield  {author} {\bibinfo {author} {\bibfnamefont {Giovanni}\
  \bibnamefont {Lerario}}, \bibinfo {author} {\bibfnamefont {Antonio}\
  \bibnamefont {Fieramosca}}, \bibinfo {author} {\bibfnamefont {Fábio}\
  \bibnamefont {Barachati}}, \bibinfo {author} {\bibfnamefont {Dario}\
  \bibnamefont {Ballarini}}, \bibinfo {author} {\bibfnamefont
  {Konstantinos~S.}\ \bibnamefont {Daskalakis}}, \bibinfo {author}
  {\bibfnamefont {Lorenzo}\ \bibnamefont {Dominici}}, \bibinfo {author}
  {\bibfnamefont {Milena}\ \bibnamefont {De~Giorgi}}, \bibinfo {author}
  {\bibfnamefont {Stefan~A.}\ \bibnamefont {Maier}}, \bibinfo {author}
  {\bibfnamefont {Giuseppe}\ \bibnamefont {Gigli}}, \bibinfo {author}
  {\bibfnamefont {Stéphane}\ \bibnamefont {Kéna-Cohen}}, \ and\ \bibinfo
  {author} {\bibfnamefont {Daniele}\ \bibnamefont {Sanvitto}},\ }\bibfield
  {title} {\enquote {\bibinfo {title} {Room-temperature superfluidity in a
  polariton condensate},}\ }\href {\doibase 10.1038/nphys4147} {\bibfield
  {journal} {\bibinfo  {journal} {Nature Phys.}\ }\textbf {\bibinfo {volume}
  {13}},\ \bibinfo {pages} {837} (\bibinfo {year} {2017})}\BibitemShut
  {NoStop}%
\bibitem [{\citenamefont {Amo}\ \emph {et~al.}(2011)\citenamefont {Amo},
  \citenamefont {Pigeon}, \citenamefont {Sanvitto}, \citenamefont {Sala},
  \citenamefont {Hivet}, \citenamefont {Carusotto}, \citenamefont {Pisanello},
  \citenamefont {Leménager}, \citenamefont {Houdré}, \citenamefont
  {Giacobino}, \citenamefont {Ciuti},\ and\ \citenamefont
  {Bramati}}]{amo_polariton_2011}%
  \BibitemOpen
  \bibfield  {author} {\bibinfo {author} {\bibfnamefont {A.}~\bibnamefont
  {Amo}}, \bibinfo {author} {\bibfnamefont {S.}~\bibnamefont {Pigeon}},
  \bibinfo {author} {\bibfnamefont {D.}~\bibnamefont {Sanvitto}}, \bibinfo
  {author} {\bibfnamefont {V.~G}\ \bibnamefont {Sala}}, \bibinfo {author}
  {\bibfnamefont {R.}~\bibnamefont {Hivet}}, \bibinfo {author} {\bibfnamefont
  {I.}~\bibnamefont {Carusotto}}, \bibinfo {author} {\bibfnamefont
  {F.}~\bibnamefont {Pisanello}}, \bibinfo {author} {\bibfnamefont
  {G.}~\bibnamefont {Leménager}}, \bibinfo {author} {\bibfnamefont
  {R.}~\bibnamefont {Houdré}}, \bibinfo {author} {\bibfnamefont
  {E.}~\bibnamefont {Giacobino}}, \bibinfo {author} {\bibfnamefont
  {C.}~\bibnamefont {Ciuti}}, \ and\ \bibinfo {author} {\bibfnamefont
  {A.}~\bibnamefont {Bramati}},\ }\bibfield  {title} {\enquote {\bibinfo
  {title} {Polariton {Superfluids} {Reveal} {Quantum} {Hydrodynamic}
  {Solitons}},}\ }\href {\doibase 10.1126/science.1202307} {\bibfield
  {journal} {\bibinfo  {journal} {Science}\ }\textbf {\bibinfo {volume}
  {332}},\ \bibinfo {pages} {1167--1170} (\bibinfo {year} {2011})}\BibitemShut
  {NoStop}%
\bibitem [{\citenamefont {Sich}\ \emph {et~al.}(2012)\citenamefont {Sich},
  \citenamefont {Krizhanovskii}, \citenamefont {Skolnick}, \citenamefont
  {Gorbach}, \citenamefont {Hartley}, \citenamefont {Skryabin}, \citenamefont
  {Cerda-Méndez}, \citenamefont {Biermann}, \citenamefont {Hey},\ and\
  \citenamefont {Santos}}]{sich_observation_2012}%
  \BibitemOpen
  \bibfield  {author} {\bibinfo {author} {\bibfnamefont {M.}~\bibnamefont
  {Sich}}, \bibinfo {author} {\bibfnamefont {D.~N.}\ \bibnamefont
  {Krizhanovskii}}, \bibinfo {author} {\bibfnamefont {M.~S.}\ \bibnamefont
  {Skolnick}}, \bibinfo {author} {\bibfnamefont {A.~V.}\ \bibnamefont
  {Gorbach}}, \bibinfo {author} {\bibfnamefont {R.}~\bibnamefont {Hartley}},
  \bibinfo {author} {\bibfnamefont {D.~V.}\ \bibnamefont {Skryabin}}, \bibinfo
  {author} {\bibfnamefont {E.~A.}\ \bibnamefont {Cerda-Méndez}}, \bibinfo
  {author} {\bibfnamefont {K.}~\bibnamefont {Biermann}}, \bibinfo {author}
  {\bibfnamefont {R.}~\bibnamefont {Hey}}, \ and\ \bibinfo {author}
  {\bibfnamefont {P.~V.}\ \bibnamefont {Santos}},\ }\bibfield  {title}
  {\enquote {\bibinfo {title} {Observation of bright polariton solitons in a
  semiconductor microcavity},}\ }\href {\doibase 10.1038/nphoton.2011.267}
  {\bibfield  {journal} {\bibinfo  {journal} {Nature Photonics}\ }\textbf
  {\bibinfo {volume} {6}},\ \bibinfo {pages} {50--55} (\bibinfo {year}
  {2012})}\BibitemShut {NoStop}%
\bibitem [{\citenamefont {Walker}\ \emph {et~al.}(2017)\citenamefont {Walker},
  \citenamefont {Tinkler}, \citenamefont {Royall}, \citenamefont {Skryabin},
  \citenamefont {Farrer}, \citenamefont {Ritchie}, \citenamefont {Skolnick},\
  and\ \citenamefont {Krizhanovskii}}]{walker_dark_2017}%
  \BibitemOpen
  \bibfield  {author} {\bibinfo {author} {\bibfnamefont {P.~M.}\ \bibnamefont
  {Walker}}, \bibinfo {author} {\bibfnamefont {L.}~\bibnamefont {Tinkler}},
  \bibinfo {author} {\bibfnamefont {B.}~\bibnamefont {Royall}}, \bibinfo
  {author} {\bibfnamefont {D.~V.}\ \bibnamefont {Skryabin}}, \bibinfo {author}
  {\bibfnamefont {I.}~\bibnamefont {Farrer}}, \bibinfo {author} {\bibfnamefont
  {D.~A.}\ \bibnamefont {Ritchie}}, \bibinfo {author} {\bibfnamefont {M.~S.}\
  \bibnamefont {Skolnick}}, \ and\ \bibinfo {author} {\bibfnamefont {D.~N.}\
  \bibnamefont {Krizhanovskii}},\ }\bibfield  {title} {\enquote {\bibinfo
  {title} {Dark {Solitons} in {High} {Velocity} {Waveguide} {Polariton}
  {Fluids}},}\ }\href {\doibase 10.1103/PhysRevLett.119.097403} {\bibfield
  {journal} {\bibinfo  {journal} {Phys. Rev. Lett.}\ }\textbf {\bibinfo
  {volume} {119}},\ \bibinfo {pages} {097403} (\bibinfo {year}
  {2017})}\BibitemShut {NoStop}%
\bibitem [{\citenamefont {Baas}\ \emph {et~al.}(2008)\citenamefont {Baas},
  \citenamefont {Lagoudakis}, \citenamefont {Richard}, \citenamefont {André},
  \citenamefont {Dang},\ and\ \citenamefont
  {Deveaud-Plédran}}]{baas_synchronized_2008}%
  \BibitemOpen
  \bibfield  {author} {\bibinfo {author} {\bibfnamefont {A.}~\bibnamefont
  {Baas}}, \bibinfo {author} {\bibfnamefont {K.~G.}\ \bibnamefont
  {Lagoudakis}}, \bibinfo {author} {\bibfnamefont {M.}~\bibnamefont {Richard}},
  \bibinfo {author} {\bibfnamefont {R.}~\bibnamefont {André}}, \bibinfo
  {author} {\bibfnamefont {Le~Si}\ \bibnamefont {Dang}}, \ and\ \bibinfo
  {author} {\bibfnamefont {B.}~\bibnamefont {Deveaud-Plédran}},\ }\bibfield
  {title} {\enquote {\bibinfo {title} {Synchronized and {Desynchronized}
  {Phases} of {Exciton}-{Polariton} {Condensates} in the {Presence} of
  {Disorder}},}\ }\href {\doibase 10.1103/PhysRevLett.100.170401} {\bibfield
  {journal} {\bibinfo  {journal} {Phys. Rev. Lett.}\ }\textbf {\bibinfo
  {volume} {100}},\ \bibinfo {pages} {170401} (\bibinfo {year}
  {2008})}\BibitemShut {NoStop}%
\bibitem [{\citenamefont {Lagoudakis}\ \emph {et~al.}(2010)\citenamefont
  {Lagoudakis}, \citenamefont {Pietka}, \citenamefont {Wouters}, \citenamefont
  {André},\ and\ \citenamefont {Deveaud-Plédran}}]{lagoudakis_coherent_2010}%
  \BibitemOpen
  \bibfield  {author} {\bibinfo {author} {\bibfnamefont {K.~G.}\ \bibnamefont
  {Lagoudakis}}, \bibinfo {author} {\bibfnamefont {B.}~\bibnamefont {Pietka}},
  \bibinfo {author} {\bibfnamefont {M.}~\bibnamefont {Wouters}}, \bibinfo
  {author} {\bibfnamefont {R.}~\bibnamefont {André}}, \ and\ \bibinfo {author}
  {\bibfnamefont {B.}~\bibnamefont {Deveaud-Plédran}},\ }\bibfield  {title}
  {\enquote {\bibinfo {title} {Coherent oscillations in an exciton-polariton
  josephson junction},}\ }\href {\doibase 10.1103/PhysRevLett.105.120403}
  {\bibfield  {journal} {\bibinfo  {journal} {Phys. Rev. Lett.}\ }\textbf
  {\bibinfo {volume} {105}},\ \bibinfo {pages} {120403} (\bibinfo {year}
  {2010})}\BibitemShut {NoStop}%
\bibitem [{\citenamefont {Wouters}(2008)}]{wouters_synchronized_2008}%
  \BibitemOpen
  \bibfield  {author} {\bibinfo {author} {\bibfnamefont {Michiel}\ \bibnamefont
  {Wouters}},\ }\bibfield  {title} {\enquote {\bibinfo {title} {Synchronized
  and desynchronized phases of coupled nonequilibrium exciton-polariton
  condensates},}\ }\href {\doibase 10.1103/PhysRevB.77.121302} {\bibfield
  {journal} {\bibinfo  {journal} {Phys. Rev. B}\ }\textbf {\bibinfo {volume}
  {77}},\ \bibinfo {pages} {121302} (\bibinfo {year} {2008})}\BibitemShut
  {NoStop}%
\bibitem [{\citenamefont {Malpuech}\ \emph {et~al.}(2007)\citenamefont
  {Malpuech}, \citenamefont {Solnyshkov}, \citenamefont {Ouerdane},
  \citenamefont {Glazov},\ and\ \citenamefont {Shelykh}}]{malpuech_bose_2007}%
  \BibitemOpen
  \bibfield  {author} {\bibinfo {author} {\bibfnamefont {G.}~\bibnamefont
  {Malpuech}}, \bibinfo {author} {\bibfnamefont {D.~D.}\ \bibnamefont
  {Solnyshkov}}, \bibinfo {author} {\bibfnamefont {H.}~\bibnamefont
  {Ouerdane}}, \bibinfo {author} {\bibfnamefont {M.~M.}\ \bibnamefont
  {Glazov}}, \ and\ \bibinfo {author} {\bibfnamefont {I.}~\bibnamefont
  {Shelykh}},\ }\bibfield  {title} {\enquote {\bibinfo {title} {Bose {Glass}
  and {Superfluid} {Phases} of {Cavity} {Polaritons}},}\ }\href {\doibase
  10.1103/PhysRevLett.98.206402} {\bibfield  {journal} {\bibinfo  {journal}
  {Phys. Rev. Lett.}\ }\textbf {\bibinfo {volume} {98}},\ \bibinfo {pages}
  {206402} (\bibinfo {year} {2007})}\BibitemShut {NoStop}%
\bibitem [{\citenamefont {Janot}\ \emph {et~al.}(2013)\citenamefont {Janot},
  \citenamefont {Hyart}, \citenamefont {Eastham},\ and\ \citenamefont
  {Rosenow}}]{janot_superfluid_2013}%
  \BibitemOpen
  \bibfield  {author} {\bibinfo {author} {\bibfnamefont {Alexander}\
  \bibnamefont {Janot}}, \bibinfo {author} {\bibfnamefont {Timo}\ \bibnamefont
  {Hyart}}, \bibinfo {author} {\bibfnamefont {Paul~R.}\ \bibnamefont
  {Eastham}}, \ and\ \bibinfo {author} {\bibfnamefont {Bernd}\ \bibnamefont
  {Rosenow}},\ }\bibfield  {title} {\enquote {\bibinfo {title} {Superfluid
  {Stiffness} of a {Driven} {Dissipative} {Condensate} with {Disorder}},}\
  }\href {\doibase 10.1103/PhysRevLett.111.230403} {\bibfield  {journal}
  {\bibinfo  {journal} {Phys. Rev. Lett.}\ }\textbf {\bibinfo {volume} {111}},\
  \bibinfo {pages} {230403} (\bibinfo {year} {2013})}\BibitemShut {NoStop}%
\bibitem [{\citenamefont {Nelsen}\ \emph {et~al.}(2013)\citenamefont {Nelsen},
  \citenamefont {Liu}, \citenamefont {Steger}, \citenamefont {Snoke},
  \citenamefont {Balili}, \citenamefont {West},\ and\ \citenamefont
  {Pfeiffer}}]{nelsen_dissipationless_2013}%
  \BibitemOpen
  \bibfield  {author} {\bibinfo {author} {\bibfnamefont {Bryan}\ \bibnamefont
  {Nelsen}}, \bibinfo {author} {\bibfnamefont {Gangqiang}\ \bibnamefont {Liu}},
  \bibinfo {author} {\bibfnamefont {Mark}\ \bibnamefont {Steger}}, \bibinfo
  {author} {\bibfnamefont {David~W.}\ \bibnamefont {Snoke}}, \bibinfo {author}
  {\bibfnamefont {Ryan}\ \bibnamefont {Balili}}, \bibinfo {author}
  {\bibfnamefont {Ken}\ \bibnamefont {West}}, \ and\ \bibinfo {author}
  {\bibfnamefont {Loren}\ \bibnamefont {Pfeiffer}},\ }\bibfield  {title}
  {\enquote {\bibinfo {title} {Dissipationless flow and sharp threshold of a
  polariton condensate with long lifetime},}\ }\href {\doibase
  10.1103/PhysRevX.3.041015} {\bibfield  {journal} {\bibinfo  {journal} {Phys.
  Rev. X}\ }\textbf {\bibinfo {volume} {3}},\ \bibinfo {pages} {041015}
  (\bibinfo {year} {2013})}\BibitemShut {NoStop}%
\bibitem [{\citenamefont {Geerligs}\ \emph {et~al.}(1989)\citenamefont
  {Geerligs}, \citenamefont {Peters}, \citenamefont {de~Groot}, \citenamefont
  {Verbruggen},\ and\ \citenamefont {Mooij}}]{geerligs_charging_1989}%
  \BibitemOpen
  \bibfield  {author} {\bibinfo {author} {\bibfnamefont {L.~J.}\ \bibnamefont
  {Geerligs}}, \bibinfo {author} {\bibfnamefont {M.}~\bibnamefont {Peters}},
  \bibinfo {author} {\bibfnamefont {L.~E.~M.}\ \bibnamefont {de~Groot}},
  \bibinfo {author} {\bibfnamefont {A.}~\bibnamefont {Verbruggen}}, \ and\
  \bibinfo {author} {\bibfnamefont {J.~E.}\ \bibnamefont {Mooij}},\ }\bibfield
  {title} {\enquote {\bibinfo {title} {Charging effects and quantum coherence
  in regular {Josephson} junction arrays},}\ }\href {\doibase
  10.1103/PhysRevLett.63.326} {\bibfield  {journal} {\bibinfo  {journal}
  {Physical Review Letters}\ }\textbf {\bibinfo {volume} {63}},\ \bibinfo
  {pages} {326--329} (\bibinfo {year} {1989})}\BibitemShut {NoStop}%
\bibitem [{\citenamefont {Fisher}\ \emph {et~al.}(1989)\citenamefont {Fisher},
  \citenamefont {Weichman}, \citenamefont {Grinstein},\ and\ \citenamefont
  {Fisher}}]{fisher_boson_1989}%
  \BibitemOpen
  \bibfield  {author} {\bibinfo {author} {\bibfnamefont {Matthew P.~A.}\
  \bibnamefont {Fisher}}, \bibinfo {author} {\bibfnamefont {Peter~B.}\
  \bibnamefont {Weichman}}, \bibinfo {author} {\bibfnamefont {G.}~\bibnamefont
  {Grinstein}}, \ and\ \bibinfo {author} {\bibfnamefont {Daniel~S.}\
  \bibnamefont {Fisher}},\ }\bibfield  {title} {\enquote {\bibinfo {title}
  {Boson localization and the superfluid-insulator transition},}\ }\href
  {\doibase 10.1103/PhysRevB.40.546} {\bibfield  {journal} {\bibinfo  {journal}
  {Phys. Rev. B}\ }\textbf {\bibinfo {volume} {40}},\ \bibinfo {pages} {546}
  (\bibinfo {year} {1989})}\BibitemShut {NoStop}%
\bibitem [{\citenamefont {Meldgin}\ \emph {et~al.}(2016)\citenamefont
  {Meldgin}, \citenamefont {Ray}, \citenamefont {Russ}, \citenamefont {Chen},
  \citenamefont {Ceperley},\ and\ \citenamefont
  {{DeMarco}}}]{meldgin_probing_2016}%
  \BibitemOpen
  \bibfield  {author} {\bibinfo {author} {\bibfnamefont {Carolyn}\ \bibnamefont
  {Meldgin}}, \bibinfo {author} {\bibfnamefont {Ushnish}\ \bibnamefont {Ray}},
  \bibinfo {author} {\bibfnamefont {Philip}\ \bibnamefont {Russ}}, \bibinfo
  {author} {\bibfnamefont {David}\ \bibnamefont {Chen}}, \bibinfo {author}
  {\bibfnamefont {David~M.}\ \bibnamefont {Ceperley}}, \ and\ \bibinfo {author}
  {\bibfnamefont {Brian}\ \bibnamefont {{DeMarco}}},\ }\bibfield  {title}
  {\enquote {\bibinfo {title} {Probing the bose glass-superfluid transition
  using quantum quenches of disorder},}\ }\href {\doibase 10.1038/nphys3695}
  {\bibfield  {journal} {\bibinfo  {journal} {Nature Phys.}\ }\textbf {\bibinfo
  {volume} {12}},\ \bibinfo {pages} {646--649} (\bibinfo {year}
  {2016})}\BibitemShut {NoStop}%
\bibitem [{\citenamefont {Cristofolini}\ \emph {et~al.}(2013)\citenamefont
  {Cristofolini}, \citenamefont {Dreismann}, \citenamefont {Christmann},
  \citenamefont {Franchetti}, \citenamefont {Berloff}, \citenamefont {Tsotsis},
  \citenamefont {Hatzopoulos}, \citenamefont {Savvidis},\ and\ \citenamefont
  {Baumberg}}]{cristofolini_optical_2013}%
  \BibitemOpen
  \bibfield  {author} {\bibinfo {author} {\bibfnamefont {P.}~\bibnamefont
  {Cristofolini}}, \bibinfo {author} {\bibfnamefont {A.}~\bibnamefont
  {Dreismann}}, \bibinfo {author} {\bibfnamefont {G.}~\bibnamefont
  {Christmann}}, \bibinfo {author} {\bibfnamefont {G.}~\bibnamefont
  {Franchetti}}, \bibinfo {author} {\bibfnamefont {N.~G.}\ \bibnamefont
  {Berloff}}, \bibinfo {author} {\bibfnamefont {P.}~\bibnamefont {Tsotsis}},
  \bibinfo {author} {\bibfnamefont {Z.}~\bibnamefont {Hatzopoulos}}, \bibinfo
  {author} {\bibfnamefont {P.~G.}\ \bibnamefont {Savvidis}}, \ and\ \bibinfo
  {author} {\bibfnamefont {J.~J.}\ \bibnamefont {Baumberg}},\ }\bibfield
  {title} {\enquote {\bibinfo {title} {Optical superfluid phase transitions and
  trapping of polariton condensates},}\ }\href {\doibase
  10.1103/PhysRevLett.110.186403} {\bibfield  {journal} {\bibinfo  {journal}
  {Phys. Rev. Lett.}\ }\textbf {\bibinfo {volume} {110}},\ \bibinfo {pages}
  {186403} (\bibinfo {year} {2013})}\BibitemShut {NoStop}%
\bibitem [{\citenamefont {Askitopoulos}\ \emph {et~al.}(2013)\citenamefont
  {Askitopoulos}, \citenamefont {Ohadi}, \citenamefont {Kavokin}, \citenamefont
  {Hatzopoulos}, \citenamefont {Savvidis},\ and\ \citenamefont
  {Lagoudakis}}]{askitopoulos_polariton_2013}%
  \BibitemOpen
  \bibfield  {author} {\bibinfo {author} {\bibfnamefont {A.}~\bibnamefont
  {Askitopoulos}}, \bibinfo {author} {\bibfnamefont {H.}~\bibnamefont {Ohadi}},
  \bibinfo {author} {\bibfnamefont {A.~V.}\ \bibnamefont {Kavokin}}, \bibinfo
  {author} {\bibfnamefont {Z.}~\bibnamefont {Hatzopoulos}}, \bibinfo {author}
  {\bibfnamefont {P.~G.}\ \bibnamefont {Savvidis}}, \ and\ \bibinfo {author}
  {\bibfnamefont {P.~G.}\ \bibnamefont {Lagoudakis}},\ }\bibfield  {title}
  {\enquote {\bibinfo {title} {Polariton condensation in an optically induced
  two-dimensional potential},}\ }\href {\doibase 10.1103/PhysRevB.88.041308}
  {\bibfield  {journal} {\bibinfo  {journal} {Phys. Rev. B}\ }\textbf {\bibinfo
  {volume} {88}},\ \bibinfo {pages} {041308} (\bibinfo {year}
  {2013})}\BibitemShut {NoStop}%
\bibitem [{\citenamefont {Ohadi}\ \emph {et~al.}(2016)\citenamefont {Ohadi},
  \citenamefont {del Valle-Inclan~Redondo}, \citenamefont {Dreismann},
  \citenamefont {Rubo}, \citenamefont {Pinsker}, \citenamefont {Tsintzos},
  \citenamefont {Hatzopoulos}, \citenamefont {Savvidis},\ and\ \citenamefont
  {Baumberg}}]{ohadi_tunable_2016}%
  \BibitemOpen
  \bibfield  {author} {\bibinfo {author} {\bibfnamefont {H.}~\bibnamefont
  {Ohadi}}, \bibinfo {author} {\bibfnamefont {Y.}~\bibnamefont {del
  Valle-Inclan~Redondo}}, \bibinfo {author} {\bibfnamefont {A.}~\bibnamefont
  {Dreismann}}, \bibinfo {author} {\bibfnamefont {Y.~G.}\ \bibnamefont {Rubo}},
  \bibinfo {author} {\bibfnamefont {F.}~\bibnamefont {Pinsker}}, \bibinfo
  {author} {\bibfnamefont {S.~I.}\ \bibnamefont {Tsintzos}}, \bibinfo {author}
  {\bibfnamefont {Z.}~\bibnamefont {Hatzopoulos}}, \bibinfo {author}
  {\bibfnamefont {P.~G.}\ \bibnamefont {Savvidis}}, \ and\ \bibinfo {author}
  {\bibfnamefont {J.~J.}\ \bibnamefont {Baumberg}},\ }\bibfield  {title}
  {\enquote {\bibinfo {title} {Tunable magnetic alignment between trapped
  exciton-polariton condensates},}\ }\href {\doibase
  10.1103/PhysRevLett.116.106403} {\bibfield  {journal} {\bibinfo  {journal}
  {Phys. Rev. Lett.}\ }\textbf {\bibinfo {volume} {116}},\ \bibinfo {pages}
  {106403} (\bibinfo {year} {2016})}\BibitemShut {NoStop}%
\bibitem [{\citenamefont {Ohadi}\ \emph {et~al.}(2017)\citenamefont {Ohadi},
  \citenamefont {Ramsay}, \citenamefont {Sigurdsson}, \citenamefont {del
  Valle-Inclan~Redondo}, \citenamefont {Tsintzos}, \citenamefont {Hatzopoulos},
  \citenamefont {Liew}, \citenamefont {Shelykh}, \citenamefont {Rubo},
  \citenamefont {Savvidis},\ and\ \citenamefont {Baumberg}}]{ohadi_spin_2017}%
  \BibitemOpen
  \bibfield  {author} {\bibinfo {author} {\bibfnamefont {H.}~\bibnamefont
  {Ohadi}}, \bibinfo {author} {\bibfnamefont {A.~J.}\ \bibnamefont {Ramsay}},
  \bibinfo {author} {\bibfnamefont {H.}~\bibnamefont {Sigurdsson}}, \bibinfo
  {author} {\bibfnamefont {Y.}~\bibnamefont {del Valle-Inclan~Redondo}},
  \bibinfo {author} {\bibfnamefont {S.~I.}\ \bibnamefont {Tsintzos}}, \bibinfo
  {author} {\bibfnamefont {Z.}~\bibnamefont {Hatzopoulos}}, \bibinfo {author}
  {\bibfnamefont {T.~C.~H.}\ \bibnamefont {Liew}}, \bibinfo {author}
  {\bibfnamefont {I.~A.}\ \bibnamefont {Shelykh}}, \bibinfo {author}
  {\bibfnamefont {Y.~G.}\ \bibnamefont {Rubo}}, \bibinfo {author}
  {\bibfnamefont {P.~G.}\ \bibnamefont {Savvidis}}, \ and\ \bibinfo {author}
  {\bibfnamefont {J.~J.}\ \bibnamefont {Baumberg}},\ }\bibfield  {title}
  {\enquote {\bibinfo {title} {Spin {Order} and {Phase} {Transitions} in
  {Chains} of {Polariton} {Condensates}},}\ }\href {\doibase
  10.1103/PhysRevLett.119.067401} {\bibfield  {journal} {\bibinfo  {journal}
  {Phys. Rev. Lett.}\ }\textbf {\bibinfo {volume} {119}},\ \bibinfo {pages}
  {067401} (\bibinfo {year} {2017})}\BibitemShut {NoStop}%
\bibitem [{\citenamefont {Ohadi}\ \emph {et~al.}(2015)\citenamefont {Ohadi},
  \citenamefont {Dreismann}, \citenamefont {Rubo}, \citenamefont {Pinsker},
  \citenamefont {del Valle-Inclan~Redondo}, \citenamefont {Tsintzos},
  \citenamefont {Hatzopoulos}, \citenamefont {Savvidis},\ and\ \citenamefont
  {Baumberg}}]{ohadi_spontaneous_2015}%
  \BibitemOpen
  \bibfield  {author} {\bibinfo {author} {\bibfnamefont {H.}~\bibnamefont
  {Ohadi}}, \bibinfo {author} {\bibfnamefont {A.}~\bibnamefont {Dreismann}},
  \bibinfo {author} {\bibfnamefont {Y.~G.}\ \bibnamefont {Rubo}}, \bibinfo
  {author} {\bibfnamefont {F.}~\bibnamefont {Pinsker}}, \bibinfo {author}
  {\bibfnamefont {Y.}~\bibnamefont {del Valle-Inclan~Redondo}}, \bibinfo
  {author} {\bibfnamefont {S.~I.}\ \bibnamefont {Tsintzos}}, \bibinfo {author}
  {\bibfnamefont {Z.}~\bibnamefont {Hatzopoulos}}, \bibinfo {author}
  {\bibfnamefont {P.~G.}\ \bibnamefont {Savvidis}}, \ and\ \bibinfo {author}
  {\bibfnamefont {J.~J.}\ \bibnamefont {Baumberg}},\ }\bibfield  {title}
  {\enquote {\bibinfo {title} {Spontaneous {Spin} {Bifurcations} and
  {Ferromagnetic} {Phase} {Transitions} in a {Spinor} {Exciton}-{Polariton}
  {Condensate}},}\ }\href {\doibase 10.1103/PhysRevX.5.031002} {\bibfield
  {journal} {\bibinfo  {journal} {Phys. Rev. X}\ }\textbf {\bibinfo {volume}
  {5}},\ \bibinfo {pages} {031002} (\bibinfo {year} {2015})}\BibitemShut
  {NoStop}%
\bibitem [{\citenamefont {Tsotsis}\ \emph {et~al.}(2012)\citenamefont
  {Tsotsis}, \citenamefont {Eldridge}, \citenamefont {Gao}, \citenamefont
  {Tsintzos}, \citenamefont {Hatzopoulos},\ and\ \citenamefont
  {Savvidis}}]{tsotsis_lasing_2012}%
  \BibitemOpen
  \bibfield  {author} {\bibinfo {author} {\bibfnamefont {P}~\bibnamefont
  {Tsotsis}}, \bibinfo {author} {\bibfnamefont {P~S}\ \bibnamefont {Eldridge}},
  \bibinfo {author} {\bibfnamefont {T}~\bibnamefont {Gao}}, \bibinfo {author}
  {\bibfnamefont {S~I}\ \bibnamefont {Tsintzos}}, \bibinfo {author}
  {\bibfnamefont {Z}~\bibnamefont {Hatzopoulos}}, \ and\ \bibinfo {author}
  {\bibfnamefont {P~G}\ \bibnamefont {Savvidis}},\ }\bibfield  {title}
  {\enquote {\bibinfo {title} {Lasing threshold doubling at the crossover from
  strong to weak coupling regime in {GaAs} microcavity},}\ }\href {\doibase
  10.1088/1367-2630/14/2/023060} {\bibfield  {journal} {\bibinfo  {journal} {N.
  J. Phys.}\ }\textbf {\bibinfo {volume} {14}},\ \bibinfo {pages} {023060}
  (\bibinfo {year} {2012})}\BibitemShut {NoStop}%
\bibitem [{Parameters()}]{Parameters}%
  \BibitemOpen
  Parameters,\ \href@noop {} {}\bibinfo {note} {$\alpha =\SI {3}{\ueV \um ^2}$;
  $g_r=2\alpha $; $g_P=0.6\alpha $; $\Lambda =0.2$; $m^*=5.1\times 10^{-5}m_e$;
  $R=\SI {0.01}{\ps ^{-1}\um ^2}$; $\gamma =\SI {0.1}{\ps ^{-1}\um ^2}$;
  $\gamma _R=62.5\gamma $;}\BibitemShut {NoStop}%
\bibitem [{\citenamefont {Krauth}\ \emph {et~al.}(1991)\citenamefont {Krauth},
  \citenamefont {Trivedi},\ and\ \citenamefont
  {Ceperley}}]{krauth_superfluid-insulator_1991}%
  \BibitemOpen
  \bibfield  {author} {\bibinfo {author} {\bibfnamefont {Werner}\ \bibnamefont
  {Krauth}}, \bibinfo {author} {\bibfnamefont {Nandini}\ \bibnamefont
  {Trivedi}}, \ and\ \bibinfo {author} {\bibfnamefont {David}\ \bibnamefont
  {Ceperley}},\ }\bibfield  {title} {\enquote {\bibinfo {title}
  {Superfluid-insulator transition in disordered boson systems},}\ }\href
  {\doibase 10.1103/PhysRevLett.67.2307} {\bibfield  {journal} {\bibinfo
  {journal} {Phys. Rev. Lett.}\ }\textbf {\bibinfo {volume} {67}},\ \bibinfo
  {pages} {2307--2310} (\bibinfo {year} {1991})}\BibitemShut {NoStop}%
\bibitem [{\citenamefont {Fallani}\ \emph {et~al.}(2007)\citenamefont
  {Fallani}, \citenamefont {Lye}, \citenamefont {Guarrera}, \citenamefont
  {Fort},\ and\ \citenamefont {Inguscio}}]{fallani_ultracold_2007}%
  \BibitemOpen
  \bibfield  {author} {\bibinfo {author} {\bibfnamefont {L.}~\bibnamefont
  {Fallani}}, \bibinfo {author} {\bibfnamefont {J.~E.}\ \bibnamefont {Lye}},
  \bibinfo {author} {\bibfnamefont {V.}~\bibnamefont {Guarrera}}, \bibinfo
  {author} {\bibfnamefont {C.}~\bibnamefont {Fort}}, \ and\ \bibinfo {author}
  {\bibfnamefont {M.}~\bibnamefont {Inguscio}},\ }\bibfield  {title} {\enquote
  {\bibinfo {title} {Ultracold atoms in a disordered crystal of light: Towards
  a bose glass},}\ }\href {\doibase 10.1103/PhysRevLett.98.130404} {\bibfield
  {journal} {\bibinfo  {journal} {Phys. Rev. Lett.}\ }\textbf {\bibinfo
  {volume} {98}},\ \bibinfo {pages} {130404} (\bibinfo {year}
  {2007})}\BibitemShut {NoStop}%
\bibitem [{\citenamefont {Winful}\ and\ \citenamefont
  {Wang}(1988)}]{winful_stability_1988}%
  \BibitemOpen
  \bibfield  {author} {\bibinfo {author} {\bibfnamefont {H.~G.}\ \bibnamefont
  {Winful}}\ and\ \bibinfo {author} {\bibfnamefont {S.~S.}\ \bibnamefont
  {Wang}},\ }\bibfield  {title} {\enquote {\bibinfo {title} {Stability of phase
  locking in coupled semiconductor laser arrays},}\ }\href {\doibase
  10.1063/1.100363} {\bibfield  {journal} {\bibinfo  {journal} {Applied Physics
  Letters}\ }\textbf {\bibinfo {volume} {53}},\ \bibinfo {pages} {1894--1896}
  (\bibinfo {year} {1988})}\BibitemShut {NoStop}%
\bibitem [{\citenamefont {Born}\ \emph {et~al.}(1999)\citenamefont {Born},
  \citenamefont {Wolf}, \citenamefont {Bhatia}, \citenamefont {Clemmow},
  \citenamefont {Gabor}, \citenamefont {Stokes}, \citenamefont {Taylor},
  \citenamefont {Wayman},\ and\ \citenamefont
  {Wilcock}}]{born_principles_1999}%
  \BibitemOpen
  \bibfield  {author} {\bibinfo {author} {\bibfnamefont {Max}\ \bibnamefont
  {Born}}, \bibinfo {author} {\bibfnamefont {Emil}\ \bibnamefont {Wolf}},
  \bibinfo {author} {\bibfnamefont {A.~B.}\ \bibnamefont {Bhatia}}, \bibinfo
  {author} {\bibfnamefont {P.~C.}\ \bibnamefont {Clemmow}}, \bibinfo {author}
  {\bibfnamefont {D.}~\bibnamefont {Gabor}}, \bibinfo {author} {\bibfnamefont
  {A.~R.}\ \bibnamefont {Stokes}}, \bibinfo {author} {\bibfnamefont {A.~M.}\
  \bibnamefont {Taylor}}, \bibinfo {author} {\bibfnamefont {P.~A.}\
  \bibnamefont {Wayman}}, \ and\ \bibinfo {author} {\bibfnamefont {W.~L.}\
  \bibnamefont {Wilcock}},\ }\href@noop {} {\emph {\bibinfo {title} {Principles
  of {Optics}: {Electromagnetic} {Theory} of {Propagation}, {Interference} and
  {Diffraction} of {Light}}}}\ (\bibinfo  {publisher} {Cambridge University
  Press},\ \bibinfo {address} {Cambridge ; New York},\ \bibinfo {year}
  {1999})\BibitemShut {NoStop}%
\end{thebibliography}%

\end{document}